\documentclass[11pt]{article}
\usepackage{graphicx}
\usepackage{amsmath, latexsym}
\usepackage{mo}

\allowdisplaybreaks

\begin{document}

\thispagestyle{empty}

\begin{flushright}
IC/2000/94 \\
hep-th/0008057
\vskip 14mm
\end{flushright}

\begin{center}
{\LARGE{ Noncommutative $\Phi^4$ Theory at Two Loops}}
\vskip.5cm

{\textbf {\Large {Andrei Micu and M.M. Sheikh-Jabbari}}}


{\it The Abdus Salam International Centre for Theoretical Physics\\
 Strada Costiera, 11. 34014, Trieste, Italy}\\
{\tt  amicu, jabbari@ictp.trieste.it }\\

\end{center}

\vskip 8mm

\begin{center}
{\textbf Abstract}
\end{center}

We study perturbative aspects of noncommutative field
theories. This work is arranged in two parts. First, we review
noncommutative
field theories in general and discuss both canonical and path integral
quantization methods. In the second part, we consider the particular
example of noncommutative $\Phi^4$ theory in four dimensions and work out
the corresponding effective action and discuss renormalizability of the
theory, up to two loops.

\parskip.025cm
\tableofcontents
\parskip.5cm




\section{Generalities}
\subsection{Introduction}

In the past two years, a lot of work has been devoted to the study of \nc{e}
field theories, i.e. field theories on the Moyal plane.
The main motivation for these theories arises from string theory:
the end points of the open strings trapped on a D-brane with a nonzero
NSNS two form B-field background turn out to be noncommuting
\cite{sh}. Then the \nc{e} field theories, in particular \nc{e}
supersymmetric (Yang-Mills) gauge theories appear as the low energy
effective theory of such D-branes \cite{{SH},{SW}}. Apart from string
theory, \nc{e} field theories are very interesting as 
field theories.  In general, when we study a field theory, we
should emphasize that it 
is ``well behaved''. From this point
of view, \nc{e} field theories are really challenging because they are
nonlocal (they involve arbitrarily high orders of the derivatives), and 
there is a dimensionful parameter, other than the masses, the \nc{ity}
parameter $\te$.
The nonlocality may
have consequences on the ``CPT theorem'' as well as the causality. On the
other hand the dimensionful parameter $\te$ may ruin the
renormalizability of the theory. It was shown in \cite{{meh},{sei2}}  that
indeed space-time \nc{ity} ($\te_{0i} \ne 0$) leads to a non-unitary
theory, while field theories on  \nc{e} space are well behaved in this respect.
However, we should add that the special case of ``light-like''
noncommutativity has also been discussed and shown that it leads to
unitary quantum theories \cite{{sei2},{Light-like}}. 

Similar to the usual field theories, one can build \nc{e} version of
scalar, Dirac and vector (gauge) theories. The \nc{e} scalar theory
with $\Phi^4$ interaction is considered in \cite{sei1}, \cite{filk},
\cite{araf}, \cite{8} and it has been shown that this theory is
renormalizable up to two loops.
Similarly, one can consider the pure \nc{e} gauge theories; in
particular \nc{e} U(N) theory has been shown to be renormalizable up
to one loop \cite{sei1}, \cite{8}, \cite{9}, \cite{armoni}. Adding
fermions to the
\nc{e} U(1) has also been studied in \cite{10}, \cite{11},
\cite{12}. However in this work we will mostly concentrate on the
scalar theory.

Usually in studying the \nc{e} field theories, the classical properties of these
theories are not addressed. At the classical level, the variational method and  the Noether theorem
are discussed in detailed. We show that the
\nc{e} version of the Noether theorem holds, i.e. the conserved current for the {\it
external} symmetries (those which are not related to space-time, like gauge
symmetry) is the usual one, but the product of the fields is now replaced by
the 
star-product.
For the internal symmetries under which the \nc{e} parameter $\theta_{mu\nu}$, is invariant (e.g. 
the translation symmetry) in general the conserved current is not conserved, i.e. its divergence
can be expressed as
Moyal 
brackets. However, we should remind that for both the internal and external
symmetries, the conserved {\it charge} is still conserved, provided that the
space-time noncommutativity ($\theta_{0i}$) is zero.
 
To quantize any given field theory there are two approaches which are usually  
equivalent \cite{wei}, the path integral and canonical quantizations.
We discuss both cases for the \nc{e} scalar theory. Along these ideas, 
this can be done for all fields, fermionic and vector fields. The
basic point is that, despite having different interactions, and because the
quadratic part of the action is the same for the commutative and \nc{e}
cases, 
the {\it perturbative} Fock space for the commutative theory and its \nc{e}
counter-part are the same. In the path integral formulation this is reflected
in the fact that the integration measures are the same for the commutative case and its \nc{e} 
version.

We then proceed with the explicit loop calculations of the real \nc{e}
$\phi^4$ theory in four dimensions. At one loop , we present the detailed calculation
of both two and four point functions. In general, for any \nc{e} field theory, the loop
diagrams can be classified in the so-called ``planar'' and ``non-planar'' graphs.  
At one loop level, the planar part of diagrams show the same UV divergence
structure as the corresponding commutative theory, while the non-planar pieces are UV
finite. 
So, altogether the counter-terms (responsible for the cancellation of UV
infinities) 
have the same structure, but with different numeric factors, compared to the
commutative counter-part. However, surprisingly it turns out that this
difference in the numeric factors is not going to destroy the (UV) renormalizability of
the \nc{e} $\phi^4$, Yang-Mills and QED theories.
As for the non-planar diagrams, we will show that, although being UV finite,
they involve IR divergences. This is what is known as UV/IR mixing, which is a
general feature of the \nc{e} field theories \cite{8}.
This UV/IR mixing can be understood intuitively: Let us suppose $x$ and
$y$ coordinates be noncommuting, $[x,y]=i\theta$. From the usual operator algebra 
one can easily conclude that $\Delta x \Delta y \geq \theta$. So, increasing
precision in $x$ direction ($\Delta x\to 0$ or UV limit) is naturally related
to the $\Delta y\to \infty$ or the IR limit.
Although in \cite{{sei1},{19}} some arguments have been presented to remove
these new IR divergences, they are not yet well-understood.

In this work, to ensure (and improve) the renormalizability issue of the \nc{e}
field theories we present the detailed calculations for the \nc{e} $\phi^4$
theory up to two loops.
We show that, as it is expected from renormalizable theory, again (UV)
infinities are canceled while we still remain with the IR divergences originating from the
non-planar diagrams. The peculiar feature which we find is the fact that
the counter-terms accounting for the UV divergences at two loops, like the
one loop case, {\it do not depend on $\theta$}.

Also we note that the \nc{e} parameter, $\theta$, will not receive any quantum
corrections up to two loops, and we expect this to remain at all loop levels
and also even non-perturbatively. From the string theory point of view this is
expected; there the B-field (which leads to noncommutativity on the D-brane worldvolume)
is one of the moduli of the theory preserved by supersymmetry. We should remind the
fact that for the gauge theories on \nc{e} torus the ``Morita equivalence'' (e.g. see Ref.
\cite{{Connes},{Hofman}}) the noncommutativity parameter is defined up to the $SL(2,Z)$
transformation \cite{{SW},{Hofman},{Hashi}}, however here wo do not discuss the compact cases. 

This work is organized as follows.
In section \ref{clas} we present some classical aspects
of \nc{e} theories deriving the equations of motion and the Noether
theorem. In section \ref{canq} we briefly discuss 
the canonical quantization procedure for \nc{e} theories. In
the next section we describe the path integral quantization which we
are going to use in section \ref{efac} to derive the two loops
expression for the effective action of the \nc{e} $\Phi^4$ theory. In
section \ref{ren} we present calculations which proves the (UV)
renormalizability of the $\Phi^4$ theory at two loops. 
The last section is devoted to remarks and conclusions.

\subsection{\Nc{} spaces}

In the usual quantum mechanics we have the well known commutation
relations:
\bea
\big[\hat X_i,\hat P_j\big] & = & i \hbar \delta_{ij} ~~ \textrm{and} \nn
\\
\big[\hat X_i,\hat X_j\big] & = & \big[\hat P_i,\hat P_j\big] =0 
\eea
However there is no evidence that at very short distances (or very
high energies) these relations should still be true. Then a natural 
generalization of above is to take the coordinates which do not
commute any more,
\bea
\label{comr}
\big[\hat X_i,\hat X_j\big] & = & i \te_{ij} \, , 
\eea
where $ \te_{ij}$ is a \emph{constant} of dimension $[L]^2$.
An immediate remark is that introducing this kind of commutation
relation between coordinates the Lorentz invariance is spoiled
explicitly. We should remember however that we assumed this feature
to appear only at very short distances, i.e. for $\te \to 0$ we should
recover the Lorentz symmetry. This is one of the main constraints on
our noncommutative field theories: at least at classical level, in the limit
$\te \to 0$ we should 
find a previously known
commutative field theory \footnote{However this in general does not imply
the reverse: the \nc{e} extension of a given theory is not unique. As
an example SO(2) and U(1) gauge theories are the same, but in \nc{e}
version they are different \cite{13}.}. In general (\ref{comr}) can be
extended to space-time coordinates:
\bea
\label{nc}
\big[\hat X_\mu,\hat X_\nu\big] & = & i \te_{\mu\nu} \, .
\eea
Here after we call a space with the above commutation relations 
as a \nc{e} space.

To construct the perturbative field theory formulation, it is more
convenient to use fields which are some functions and not operator
valued objects. To pass to such fields while keeping (\ref{nc})
property one should redefine the multiplication law of functional
(field) space. This new multiplication is induced from (\ref{nc})
through the so called Weyl-Moyal correspondence \cite{15}:
$$
\hat \Phi(\hat X) \longleftrightarrow \Phi(x)\ ;
$$
\bea
\hat \Phi(\hat X) & = & \int_\alpha e^{i \alpha \hat X} \;
\phi(\alpha) \; d\alpha \nn \\
\phi(\alpha) & = & \int e^{-i \alpha x}~ \Phi(x) \; dx \, , 
\eea
where $\alpha$ and $x$ are real variables. Then,
\bea
\hat \Phi_1(\hat X)\; \hat \Phi_2(\hat X) & = & \iint_{\alpha\; \beta}
e^{i \alpha \hat X} \; \phi(\alpha) \; e^{i \beta \hat X} \;
\phi(\beta) \; d\alpha \; d\beta \nn \\*
& = & \iint_{\alpha\; \beta} e^{i (\alpha+ \beta) \hat X -
\frac{1}{2} \alpha_\mu \beta_\nu [\hat X_\mu,\hat X_\nu]}
\; \phi_1(\alpha) \; \phi_2(\beta) \; d\alpha \; d\beta\ ,
\eea
and hence, 
\bea\label{defst}
\hat \Phi_1(\hat X)\; \hat \Phi_2(\hat X) \longleftrightarrow 
\Big(\Phi_1 \star \Phi_2\Big)(x)\ ,
\eea
\bea 
\Big(\Phi_1 \star \Phi_2\Big)(x) \equiv
\bigg[\st{\xi}{\eta} \; \Phi_1(x+\xi) \;\Phi_2(x+\eta) 
\bigg]_{\xi= \eta = 0}\ .
\eea
This suggests that we can work on a usual commutative space for which
the multiplication operation is modified to the so called star
product (1.7). It is easy to check that the Moyal bracket (the
commutator in which the product is modified with a star product) of
two coordinates $x_\mu$ and $x_\nu$ gives exactly the desired
commutation relations, (\ref{nc})
\bea
[x_\mu, x_\nu]_{MB} & = & i\; \te_{\mu \nu}
\eea

\subsection{Properties of the star product}

Here we summarize some useful identities of the star product algebra.

\begin{enumerate}
\item The star product between exponentials: 
\bea
\label{pr1}
e^{i k x} \star e^{i q x} & = & e^{i (k+q) x} ~\kstm
{k}{q} \, , \textrm{where} \nn \\
k \te p & \equiv & k^\mu p^\nu \te_{\mu \nu}
\eea

\item Momentum space representation:

Let $\tilde f(k)$ and $\tilde g(k)$ be the Fourier components of $f$
and $g$. Then using (\ref{pr1})
\bea
\label{pr2}
\big(f\star g \big)(x) & = & \int \db^4 k ~ \db^4 q ~\tilde f(k) ~\tilde
g(q) ~ \kstm{k}{q} ~e^{i (k+q) x} \, .
\eea

\item Associativity:
\bea
\label{pr3}
\Big[\big(f\star g\big) \star h \Big](x) & = &\Big[f\star \big( g
\star h \big) \Big](x) \, ,
\eea
which can be proved immediately if we go to momentum space.
\bea
\textrm{rhs} & = & \int \db^4 k ~ \db^4 q ~\db^4 p ~ \tilde f(k) ~
\tilde g(q) ~ \tilde h(p) ~ \kstm{k}{q} ~ \kstm{(k+q)}{p} ~ e^{i
(k+q+p) x} \, , ~~~~\textrm{and} \nn \\
\textrm{lhs} & = & \int \db^4 k ~\db^4 q ~\db^4 p ~\tilde f(k) ~ \tilde
g(q) ~\tilde h(p) ~\kstm{q}{p} ~ \kstm{k}{(q+p)} ~ e^{i (k+q+p) x} \, .
\eea

\item Star products under integral sign
\bea
\label{pr4}
\int \big(f\star g\big)(x) \; d^4 x & = & \int \big(g\star f\big)(x)
\; d^4 x  ~ = ~ \int \big( f\cdot g \big) (x) \; d^4 x \, .
\eea
Using (\ref{pr2}) we can immediately perform the integration over $x$
which will give a $\delta^4(k+q)$. 
Due to the antisymmetry of $\te$ the exponent vanishes and so:
\bea
\int \big(f\star g\big)(x) \; d^4 x & = & \int \db^4 k ~\tilde
f(k) \tilde g(-k) \nn \\*
& = & \int \big( f\cdot g\big) (x) \;d^4 x 
\eea

From (\ref{pr4}) we can deduce the cyclic property:
\bea
\label{pr5}
\int\big( f_1 \star f_2 \star \ldots f_n \big)(x) ~d^4x & = &
\int\big( f_n \star f_1 \star \ldots f_{n-1} \big)(x) ~d^4x \, .
\eea

\item Complex conjugation.

\bea 
\label{pr6}
\big( f\star g)^\ast & = & g^\ast \star f^\ast \, .
\eea
It is obvious that if $f$ is a real function then $f\star f$ is also
real. 


\end{enumerate}

\section{\Nc{} field theory at classical level}
\label{clas}

As we have seen in the previous section the way to treat the \nc{e}
theories is to modify the usual product of fields with the star
product. So, for example, the action for the \nc{e} analog of the real
$\Phi^4$ theory will be:
\bea
S[\Phi] & = & \int d^4x\left[\frac{1}{2} \partial_\mu \Phi \star
\partial^\mu \Phi - \frac{m^2}{2} \Phi \star \Phi - \frac{\l}{4!} \Phi
\star \Phi \star \Phi \star \Phi \right]
\eea

Thanks to (\ref{pr4}), the quadratic part of the action is
the same as the commutative case. Therefore the only thing which is
modified is the interaction. This is a very important point to keep in
mind that the free theory is \emph{the same} as in the commutative
case. 
However we should remind that this is not true for the topologically
non-trivial 
spaces \cite{chaich}.

\subsection{Conjugate momentum and equations of motion}

The classical equations of motion, similar to the commutative case, are
obtained by minimizing the action, i.e.
\bea
\frac{\delta S}{\delta \Phi} & = & 0 \, .
\eea
The right meaning of the functional derivatives can be found in the Appendix
\ref{fd}. Using these results we can write the equation of motion for the
scalar filed theory with a $\Phi^4$ interaction:
\bea
(\Box +m^2) \Phi & = & - \frac{\l}{3!}~\big(\Phi \star \Phi \star 
\Phi\big)(x)
\eea

In order to find the conjugate momentum we should first distinguish
two major cases:
\begin{itemize}
\item $\te_{0i} =0\ ,$
\item $\te_{0i} \ne 0\ .$
\end{itemize}

\underline{$\te_{0i}=0$}

In this case the only place where we encounter time derivatives is the
kinetic term so the conjugate momentum is the same as in the
commutative case.

\underline{$\te_{0i} \ne 0$}

This case is more delicate since we have infinite number of time
derivatives in the interaction term. It is obvious right from the
beginning that there is something non-trivial in this case; the conjugate
momentum depends 
on the interaction terms. The infinite number of time derivatives suggests
us that the theory is nonlocal in time so causality may be violated
\cite{sei2}. It was also shown that at quantum level
unitarity is not preserved any more \cite{meh}
\footnote{The case of $\te_{0i} \ne 0$ for a cylinder has recently
been discussed in \cite{16}.}. For these reasons we
will restrict ourselves only to the case with $\te_{0i}=0$ from now on.

\subsection{Noether Theorem}

Now that we have developed the functional differentiation we can
extend the Noether theorem to the \nc{e} field theories. Suppose our
action has a \emph{global continuous} symmetry. For an infinitesimal
transformation  we can write: 
\bea
S[\Phi] & = & S[\Phi +\varepsilon\; \mathcal{F}(\Phi)] \, ,
~~\textrm{with}~~ \varepsilon = \textrm{constant} \, .
\eea
Taking now an $x$-dependent $\varepsilon$ we define the current $J$
through the relation:
\bea
\label{J}
S[\Phi + \varepsilon(x) \; \mathcal{F}] - S[\Phi] & \equiv & -\int
J^\mu\big(\Phi(x) \big)\; \partial_\mu\, \varepsilon(x)
\eea

By definition the action is stationary for \emph{any} field
variation around the classical path i.e. $\frac{\delta S}{\delta \Phi}
=0$. In particular for $\delta \Phi = \varepsilon(x)\;\mathcal{F}$ 
eq. (\ref{J}) becomes:
\bea
\left. \int J^\mu\big(\Phi(x) \big)\; \partial_\mu\, \varepsilon(x)
\right|_{\substack{\textrm{classical} \\ \textrm{path}}} & = & 0 \, .
\eea
Integrating by parts we find:
\bea
\label{cons}
\int \partial_\mu J^\mu\big(\Phi(x) \big) \, \varepsilon(x) \; d^4x &
= & 0\, ,
\eea
for any $\varepsilon(x)$. So the current $J$ is conserved. This result
is very general and it can be applied for any kind of \nc{e} theory.
The notion of conserved current is a little different from the
commutative case. Due to the property (\ref{pr4})
\bea
\int [ f,g ]_{MB}\; d^4x & = & 0
\eea
so the most we can say from eq. (\ref{cons}) is:
\bea
\label{consnc}
\partial_\mu \; J^\mu & = & [ f,g ]_{MB}\ ,
\eea
for some proper functions $f,\ g$. 
This result is somehow natural since in the limit $\te \to 0$ the Moyal
bracket vanishes and we recover the classical result $\partial_\mu \;
J^\mu = 0 $.

Let us see now what happens to the charge which in the commutative
case was conserved
\bea
Q & = & \int J^0 \; d^3 x\ .
\eea
Since we are considering only the case $\te_{0i} =0$, we can repeat
the argument we have used to prove (\ref{pr4}) for the case of 
integration only over the space coordinates and we conclude
\bea
\int [f,g]_{\substack{MB \\ \te_{0i}=0}} d^3 x & = & 0
\eea
This means that if we integrate (\ref{consnc}) over the spatial coordinates we
get:
\bea
\partial_0 \int J^0 \; d^3 x + \int \vec{\nabla} \cdot \vec{J} d^3 x & = &
0
\eea
and from here we can say that, as in the commutative case, the charge
$Q$ is conserved. Note that this is true only for $\theta_{0i}=0$ and
for $\theta_{0i}\neq 0$ even the notion of the conserved charge is
ill-defined.



For \emph{external} (space-time) symmetries, e.g. translations, 
under which the noncommutativity parameter, $\theta_{\mu\nu}$,   
remains invariant, one can
also work out the corresponding conserved current. For clarity, let us
consider this particular case:
\bea
\Phi & \longrightarrow & \Phi +\delta \Phi \, , \nn \\
\delta \Phi & = & \varepsilon^\mu \partial_\mu \Phi \, , \nn \\
x_\mu & \longrightarrow & x_\mu +\varepsilon_\mu
\eea
For the action of the form:
\bea
S & = & \int d^4 x \; \ml(\Phi, \partial \, \Phi)
\eea
where
\bea
\ml & = & \frac{1}{2} \left( \partial_\mu \Phi \star
\partial^\mu \Phi - m^2 \Phi \star \Phi \right) + V_\star(\Phi)
\eea
we find:
\bea
\left. \delta S \right|_{\delta \Phi = \varepsilon^\mu  \partial_\mu
\Phi} & = & \int d^4 x \; \left[ \frac{1}{2} \partial_\mu \left( 
\partial^\mu \Phi \star \partial_\nu \Phi \; \varepsilon^\nu +
\varepsilon^\nu \, \partial_\nu \Phi \star \partial^\mu \Phi \right) -
\partial_\mu \left( \varepsilon^\mu \ml \right) \right]\ .
\eea
If we take $\Phi$ to be the classical path, i.e. $\delta S =0$ we can
write:
\bea
\int \partial_\mu \left( T_{\mu\nu} \right) \; \varepsilon^\nu \; d^4
x & = & 0 \, ,
\eea
where
\bea
T_{\mu\nu} & = & \frac{1}{2} \left( \partial_\mu \Phi \star
\partial_\nu \Phi + \partial_\nu \Phi \star \partial_\mu \Phi\right) -
g_{\mu\nu} \ml \ .
\eea

However we should remind that the divergence of $T_{\mu\nu}$ is not
zero, e.g. for the particular case of $V_\star (\Phi) = \frac{\l}{4!}
\Phi^{\star 4} $ we can write:
\bea
\partial_\mu T^{\mu\nu} & = & \frac{1}{2} \Big[\Box \; \Phi \star
\partial^\nu \Phi + \partial^\mu \Phi \star \partial_\mu \partial^\nu
\Phi + \partial_\mu \partial^\nu \Phi \star \partial^\mu \Phi +
\partial^\nu \Phi \star \Box \; \Phi \big] \nn \\
& & - \frac{1}{2} \Big[\partial^\nu \partial_\mu \Phi \star \partial^\mu
\Phi + \partial_\mu \Phi \star \partial^\mu \partial^\nu \Phi \Big] +
\frac{m^2}{2} \Big[\partial^\nu \Phi \star \Phi -  \Phi \star
\partial^\nu \Phi \Big] \nn \\
& & + \frac{\l}{4!} \Big[ \partial^\nu \Phi \star \Phi^{\star 3} +
\Phi \star \partial^\nu \Phi \star \Phi^{\star 2} + \Phi^{\star 2}
\star \partial^\nu \Phi \star \Phi + \Phi^{\star 3} \star \partial^\nu
\Phi \Big] \nn \\
\eea
Using the equations of motion for the $\Phi^4$ case:
\bea
\partial_\mu \partial^\mu \Phi + m^2 \Phi + \frac{\l}{3!} \Phi^{\star
3} & = & 0  
\eea
we can rewrite the divergence of the energy-momentum tensor 
\bea
\partial_\mu T^{\mu\nu} & = & -\frac{\l}{2\cdot 3!} \Big[ \Phi^{\star
3} \star \partial^\nu \Phi + \partial^\nu \Phi \star \Phi^{\star 3}
\Big] \nn \\
& & + \frac{\l}{4!}\Big[ \partial^\nu \Phi \star \Phi^{\star 3} +
\Phi \star \partial^\nu \Phi \star \Phi^{\star 2} + \Phi^{\star 2}
\star \partial^\nu \Phi \star \Phi + \Phi^{\star 3} \star \partial^\nu
\Phi \Big] \nn \\
& = & \frac{\l}{4!} \Big[ -\partial^\nu \Phi \star \Phi^{\star 3} +
\Phi \star \partial^\nu \Phi \star \Phi^{\star 2} + \Phi^{\star 2}
\star \partial^\nu \Phi \star \Phi - \Phi^{\star 3} \star \partial^\nu
\Phi \Big] \nn \\
& = & \frac{\l}{4!} \Big[ [\Phi,\partial^\nu \Phi]_{MB} \star
\Phi^{\star 2} - \Phi{\star 2} \star [\Phi, \partial^\nu \Phi]_{MB} 
\Big] \nn \\
& = & \frac{\l}{4!} \left[\left[\Phi,\partial^\nu \Phi \right]_{MB},
\Phi^{\star 2} \right]_{MB}
\eea
which, of course, along the earlier discussions on the conserved
charges is not going to destroy the energy-momentum conservation, for
$\theta_{0i}=0$ cases. 


If it happens that the Lagrangian density is invariant under
some \emph{internal} symmetry, we can compute explicitly the Noether
current.

For this we assume that the Lagrangian, as in the commutative
theory, depends only on $\Phi$ and $\partial \; \Phi$ and we will make
abstraction of the internal structure of the star product. It is well
known that when we vary the Lagrangian we find some terms which yield
the equation of motion in the Lagrangian representation, and
also a surface term which will give the Noether current. This
surface terms can only come from the kinetic part of the Lagrangian. 
For a term of the form
\bea
\ml_{kin} \big(\Phi(x), \partial \; \Phi(x)\big) & = & \mathcal{F}^\mu
\big(\Phi(x), \partial \; \Phi(x)\big) \star \partial_\mu \Phi \, ,
\eea
the corresponding surface term will appear when we vary the
$\partial_\mu \Phi$ and the part which enters the conserved current
corresponding to this variation is:
\bea
\mathcal{J}_\mu & = & \mathcal{F}_\mu \star \delta \Phi
\eea

Let us consider as a first example a theory with fermions, i.e. QED, 
\bea
\ml_{kin} \big(\Psi,\bar \Psi) & = & \bar \Psi \star \left(i
\gamma_\mu \partial_\mu \right) \Psi \textrm{ with the symmetry:} 
\nn \\ 
\Psi & \to & e^{i\alpha} \; \Psi ~~~~\textrm{and} \nn \\
\bar\Psi & \to & e^{-i \alpha} \; \bar \Psi \, .
\eea

Taking $\Psi$ to $\Psi + \delta\Psi$ we can write:
\bea
\delta \ml_{kin} & = & \bar \Psi \star \big(i
\gamma_\mu \partial_\mu \big) \big(\Psi + \delta \Psi\big) - \bar \Psi
\star \big(i \gamma_\mu \partial_\mu \big) \Psi  \nn \\
& = & \bar \Psi \star \big(i \gamma_\mu \partial_\mu \big)
\delta \Psi \nn \\
& = & \partial_\mu \Big(\Psi \star \big(i \gamma_\mu
\partial_\mu \big) \delta \Psi\Big) - \big(\partial_\mu \bar\Psi
\big) \star \big(i \gamma^\mu \delta\Psi\big) \, .
\eea
For an infinitesimal symmetry transformation $\delta \Psi$ will be:
\bea
\delta \Psi & = & \varepsilon \Psi \, ,
\eea
so that for a global symmetry the current takes the form:
\bea
\mathcal{J}_\mu & = & i \bar \Psi \star \big(\gamma_\mu \psi\big) \,
.
\eea

For the case of local symmetry, we can encounter two types of fermions
(say type $a$ and $b$) and consequently two different symmetry
transformations \cite{10},
\cite{11}. The arguments we have presented up to now are still valid
for a local symmetry so that the conserved currents will be:
\begin{subequations}
\bea
\mathcal{J}^a_\mu & = & \bar \Psi \gamma_\mu \star \Psi \star
\varepsilon \\
\mathcal{J}^b_\mu & = & \bar \Psi \gamma_\mu \star \varepsilon \star
\Psi 
\eea
\end{subequations}


\allowdisplaybreaks


\section{Canonical quantization of \nc{e} theories}
\label{canq}

\subsection{Scalar theory}

Here we consider scalar theories with arbitrary interaction
$V_\star (\Phi)$. The star means that the interaction contains terms
with star products, however the precise form of this terms is not
important for the general discussion. Let $S$ be the action of our
theory: 
\bea
S & = & \int d^4 x\left[\frac{1}{2} \partial_\mu \Phi \partial^\mu
\Phi - \frac{m^2}{2} \Phi^2 - V_\star (\Phi) \right] \, .
\eea

Since the free part of the action is identical to the one in the
commutative case, it is convenient to choose the Fock space and
in particular the vacuum state \emph{to be exactly the same} as in the
corresponding commutative theory so, the fields can be expanded in
terms of the same (compared to the commutative case) creation and
annihilation operators 
\bea
\Phi(x) & = & \sum_k \left[ a(k) e^{-i k x} + a^\dagger(k)
e^{i k x} \right]\ e^{i\omega t} \, .
\eea

For applying the canonical quantization method we should first compute
the conjugate momenta $\Pi(x)$ and then impose the quantization conditions
\bea
\label{qcps}
\left[\Phi(\vec x, t),\Pi(\vec y,t)\right] & = & i \delta^{(3)} (\vec
x - \vec y) \, .
\eea

However a naive application of this method may lead to severe
problems. First as we noticed for the classical theory, in the case
$\te_{0i} \ne 0$ the theory seems to be problematic \cite{meh},
\cite{17}.  That is why  we study only the case $\te_{0i} = 0$. For
this case the conjugate momentum is just the usual one which appears
in the commutative theory:
\bea 
\Pi = 
\partial_0 \Phi \, .
\eea

In the commutative case position and momentum space are completely
equivalent and we can perform our quantization where we like. However,
in the \nc{e} theory there is an ambiguity in applying the quantization
conditions in the position space. In general we know that in order to
deal with a \nc{e} space we should work in a usual space and we should
replace the products between functions with the star product. But
the quantization conditions (\ref{qcps}) are defined for $\Phi$ and
$\Pi$ computed in different points, while the star product makes sense
only between functions computed in the same point (see
eq. (\ref{defst})). We can escape these problems, if from the
very beginning we work in the momentum space and apply directly the
quantization conditions in the momentum space:
\bea
\label{qcms}
\left[\tilde\Phi(k),\tilde\Pi(q)\right] & = & i \delta^{(4)}(k-q)
\, . 
\eea

This is possible because in momentum space the difference between the
usual commutator and the Moyal bracket is just a phase factor
$e^{i k\te q}$ which has no relevance due to the
$\delta$-function which appears in the rhs of eq. (\ref{qcms}).

From this point the quantization can go on in the same way as in the
commutative case. At the level of the free theory everything is the
same and only the interaction keeps track of the \nc{e} structure of
the space through the star product. 

\subsection{Fermionic theories}

For fermions we can apply the same arguments as in the previous
section. The free action for fermions reads:
\bea
S_{free} & = & \int d^4 x \; \Bar\Psi \Big( i \gamma^\mu \partial_\mu
-m \Big) \Psi \, ,
\eea
where $\Psi(x)$  and $\bar\Psi(x)$ can be expanded in Fourier modes:
\bea
\Psi(x) & = & \sum_k \left[ b(k) \; u(k) \;  e^{-i k x} +
d^\dagger (k) \; v(k) \; e^{i k x} \right]\ e^{i\omega t}\ ~~\textrm{and} \nn \\
\bar \Psi(x) & = & \sum_k \left[ d(k) \, u^\dagger (k) \, e^{-i k x} +
b^\dagger (k) \, v^\dagger (k) \, e^{i k x} \right]\ e^{i\omega t} \, .
\eea

As for the scalar field theory, quantization in position space is
ambiguous so we are going to use directly 
the quantization conditions in momentum space:
\bea
\big\{ \psi(k) , \psi^\dagger(q) \big \} & = &
\delta^{(4)} (k-q) \, .
\eea

For the gauge fields, because of the gauge fixing problem,  the
canonical quantization is more non-trivial, however they are out of the scope of the present work.
The only comment we are going to make is that for the gauge theories
where the canonical quantization works in the commutative case, the
procedure similarly goes through for the \nc{e} case.

\subsection{Interactions}

The next step is to introduce an interaction and derive the Feynman
rules. For simplicity we shall restrict ourselves to the scalar theory
with $\Phi^4$ interaction, but the arguments can be applied in the
same way for other theories.

Let $\phi(k)$ be the Fourier components of $\Phi\,$:
\bea
\Phi(x) & = & \int \db^4 k~ e^{i kx} \phi(k)
\eea
Then:
\bea
S_{int} & = & \frac{\l}{4!} \int d^4 x ~\Phi\star \Phi\star \Phi\star
\Phi \nn \\*
& = & \frac{\l}{4!} \int d^4 x ~\left(\Phi \star \Phi\right) \cdot 
\left( \Phi\star \Phi \right)  \\*
& = & \frac{\l}{4!} \int \db^4 x \int \db^4 k_1 \ldots \db^4 k_4 \;
\kstm{k_1}{k_2} \:\kstm{k_3}{k_4} \:e^{i \left(k_1 + k_2 + k_3 + k_4
\right)\,x} \;\phi(k_1) \; \phi(k_2) \; \phi(k_3) \; \phi(k_4) \nn \\ 
& = & \frac{\l}{4!} \int \db^4 k_1 \ldots \db^4 k_4~
\kstm{k_1}{k_2}~ \kstm{k_3}{k_4} ~\phi(k_1) \; \phi(k_2) \; \phi(k_3)
\; \phi(k_4) \times \nn \\*
& & \hspace{8cm} \times (2\pi)^4 \delta^{(4)} (k_1 + k_2 + k_3 + k_4)
\, .\nn  
\eea

Except the exponential, all the factors are symmetric in $k_1 \ldots
k_4$, hence we get:
\bea
S_{int} & = & \frac{\l}{3\cdot 4!} \int \db^4 k_1 \ldots \db^4 k_4~ 
\phi(k_1) \; \phi(k_2) \; \phi(k_3) \; \phi(k_4) \; (2\pi)^4
\delta^{(4)} (k_1 + k_2 + k_3 + k_4) \times \nn \\
& \times & \left[ \c{k_1}{k_2} \c{k_3}{k_4} + \c{k_1}{k_3}
\c{k_2}{k_4} + \c{k_1}{k_4} \c{k_2}{k_3} \right] \, .
\eea

Therefore the only difference which appears in the \nc{e} theory,
compared to the commutative one, is that for every vertex in \nc{e}
$\Phi^4$ theory we should  multiply by an additional factor: 
\bea
\label{factv}
\frac{1}{3} \left[ \c{k_1}{k_2} \c{k_3}{k_4} + \c{k_1}{k_3}
\c{k_2}{k_4} + \c{k_1}{k_4} \c{k_2}{k_3} \right] \nn \\
\eea

\section{Path integral quantization of \nc{e} theories}
\label{path}

In this section we develop the path integral formulation for
\nc{e} field theories. Although we specialize to the scalar
$\Phi^4$ theory, our method and all the arguments are valid for any
\nc{e} field theory. Then we give the diagrammatic expression of the
effective action up to two loops which we are going to use in the
study of the renormalizability of this theory in section \ref{ren}.

\subsection{Measures}

For the path integral formalism we should modify the measure of the
functional integral according to the convention we are using, i.e. to
replace the usual products between functions with the star product
\bea
\left( D \Phi \star\right) & = & \lim_{N\to \infty} d\Phi(x_1) \star
d\Phi(x_2) \star \ldots \star d\Phi(x_n) \, .
\eea

However in momentum space the star product just introduces a phase
factor which in general is going to disappear when we impose the
normalization condition for the partition function. So, as far as the
measure is concerned, we can forget about the \nc{e} structure of the
space and work with the usual measure. The fact that the measures in
the \nc{e} case should be the same as the commutative ones, in the
canonical quantization method translates into the point that the
\emph{perturbative} Hilbert space for these theories are the same.
The same kind of arguments hold for other theories as fermions and
gauge fields \footnote{However for the gauge fields one should note
that the ``physical'' measure in which ghosts (or gauge degrees of
freedom) have been thrown away, is the same for \nc{e} and commutative
cases. This in turn provides a strong support for the so called
Seiberg-Witten map \cite{SW} between commutative and \nc{e}
theories.}. In what we are interested, \emph{the perturbation theory}, we
can consider that the measure remains unchanged.


\subsection{N-point functions and effective action for \nc{e}
theories}

As we emphasized in the previous section, the \nc{e} free theory is
the same as the commutative one. The only thing that is changed is the
interaction.  So, we can say that we are dealing with a usual field
theory defined on a usual space with the usual Hilbert space, but
with strange interactions. For this reason, the \nc{e} correlation
functions should be defined in the same way as in the commutative
theory. 
\bea
G^{(n)}(x_1 \ldots x_n) & = & \left<0 \left|T \left(\hat{\Phi}(x_1)
\ldots \hat{\Phi}(x_n) \right) \right| 0 \right> \, . 
\eea  
From which we can conclude that the partition function has the same
form as in the usual case, 
\bea
Z[J] & = & \int \left(D\Phi\right) ~ e^{i S_{nc} + i \int
d^4x J(x) \Phi (x)} \, ,
\eea
and from here the generating functional for connected
graphs
\bea
Z[J] & = & e^{i W[J]} \, ,
\eea
and the effective action
\bea
\G[\Phi_c] & = & W[J] - \int  J(x) \; \Phi_c(x)\; d^4x \qquad \textrm{where}
\nn \\* 
\Phi_c(x) & = & \frac{\delta W[J]}{\delta J(x)} \, .
\eea

At this point we can repeat the calculation from the commutative case
in order to find an analytic expression for the effective action:
\bea
\label{sef1}
0 & = & \int \left(D\Phi\right) \left(\frac{\hbar}{i}\right)\cdot
\frac{\delta}{\delta \Phi(x)} ~e^{\frac{i}{\hbar} \left[
S_{nc}(\Phi) + \int J(y)\Phi(y)\, dy \right]} \nn \\
& = & \int \left(D\Phi\right)\left( \frac{\delta S}{\delta \Phi(x)} +
J(x) \right) ~e^{\frac{i}{\hbar} \left[ S_{nc}(\Phi) + \int
J(y)\Phi(y)\, dy \right]} \, .
\eea

To perform the functional integral we should
replace, as in the commutative case, $\Phi(x)$ with
$\frac{\hbar}{i} \, \frac{\delta}{\delta J(x)}\,$. However because of
the star products which appear in $ \frac{\delta S}{\delta J(x)}\,$,
in the \nc{e} case this replacement requires more attention.
In the following we are going to explain this step in detail for the
case of the scalar $\Phi^4$ theory. The only term in $ \frac{\delta
S}{\delta J}$ which still contains star products is
\bea
\frac{\delta S_{int}(\Phi)}{\delta \Phi(x)} & = & - \frac{\l}{6}
\big( \Phi \star \Phi \star \Phi \big)(x) \, .
\eea
The star product can be expanded in terms with infinite number of
partial derivatives, and taking into account that the partial and 
the functional derivatives commute, we have:
\bea
& & \hspace{-.5cm}\int \left(D\Phi\right) \big( \Phi \star\Phi \star
\Phi \big)(x) ~ e^{\frac{i}{\hbar} \left[ S_{nc}(\Phi) + \int
J(y)\Phi(y)\, dy \right]} = \nn \\
& = & \bigg[ \st{\xi}{\eta}~\str{\alpha}{\beta} \int \left(D\Phi\right) 
\Phi(x+\xi) \; \Phi(x+\eta +\alpha) \; \Phi(x+\eta +\beta) \times \nn \\
& & \hspace{4cm}
e^{\frac{i}{\hbar} \left[ S_{nc}(\Phi) + \int J(y)\Phi(y)\,
dy \right]} \bigg]_{\{\xi\}=0} = \nn \\
& = & \bigg[ \st{\xi}{\eta}~\str{\alpha}{\beta} \left(\frac{\hbar}{i}
\right)^3 \, \frac{\delta}{\delta J(x+\xi)} \frac{\delta}{\delta
J(x+\eta + \alpha)} \frac{\delta}{\delta J(x+\eta + \beta)}~
e^{\frac{i}{\hbar} W[J]} \bigg]_{\{\xi\}=0} \equiv
\nn \\ 
\nn \\
& & \hspace{4cm}\equiv \left(\frac{\hbar}{i} \right)^3 ~
\left(\frac{\delta}{\delta J} \star \frac{\delta}{\delta J} \star
\frac{\delta}{\delta J} \right)(x)~ e^{\frac{i}{\hbar} W[J]} \, ,
\eea
where the notation $\{\xi\}=0$ means $\xi=\eta=\alpha=\beta=0$.
This is a formal way of writing this result in order to make the
resemblance with the commutative case more clear, but it is not
completely wrong since the functional derivative $\frac{\delta
F[J]}{\delta J(x)}$ of a functional $F$ \emph{is a function of $x$}.
With this remark we can write the effective action (for any theory) as
in the commutative case:
\bea
\label{smth}
\frac{\delta \G[\Phi_c]}{\delta \Phi_c(x)} & = & \left(\frac{\delta
S}{\delta \Phi(x)}\right)_{\Phi(x) \to \Phi_c(x) + \frac{\hbar}{i}
\int G(x,x')\frac{\delta}{\delta \Phi_c(x')}\, d^4x'} \, .
\eea

\section{The effective action for the \nc{e} $\Phi^4$ theory}
\label{efac}

As in the usual field theories, we study the effective action and the
Green's (two point) function through the power expansion in $\hbar$:
\bea 
\G[\Phi_c] & = & \G_0(\Phi_c) + \frac{\hbar}{i} \, \G_1(\Phi_c) +
\left(\frac{\hbar}{i} \right)^2 \G_2(\Phi_c) + \ldots \nn \\
\\
G^{ij} & = & G_0^{ij} + \frac{\hbar}{i} G_1^{ij} +
\left(\frac{\hbar}{i} \right)^2 G_2^{ij} + \ldots \nn \, ,
\eea
where $\G_i$ and $G_i$ are the i-th order loop corrections.

\subsection{One loop effective action}

Using (\ref{smth}) in this subsection we work out the one-loop
effective action. As we stressed before, the only thing which is
different from the commutative case is the interaction term, so we
only consider this term. 
\begin{subequations}
\bea
\left( \frac{\delta S_{int}}{\delta \Phi(x)} \right)_{\Phi \to \Phi_c
+ \frac{\hbar}{i} \int G \frac{\delta}{\delta \Phi_c}} & = & -
\frac{\l}{6} ~e^{i W[J]} \left(\Phi_c(x) + \frac{\hbar}{i} \int d^4y
\; G(x,y) \frac{\delta}{\delta \Phi_c(y)} \right) \star \nn \\
& & \star \left(\Phi_c(x) + \frac{\hbar}{i} \int d^4y \; G(x,y)
\frac{\delta}{\delta \Phi_c(y)} \right) \star \Phi_c(x) = \nn \\
& = & - \frac{\l}{6} ~e^{i W[J]} \Big( \Phi_c \star\Phi_c \star
\Phi_c \Big)(x) \nn \\
\label{suba}
& & - \frac{\l}{6}\,\frac{\hbar}{i}~ e^{i W[J]}
\Phi_c(x) \star \int d^4y \; G(x,y) \frac{\delta}{\delta \Phi_c(y)}
\star \Phi_c(x)  \\ 
\label{subb}
& & -\frac{\l}{6} \, \frac{\hbar}{i}~ e^{i W[J]} \int d^4y
\; G(x,y) \frac{\delta}{\delta \Phi_c(y)} \star \big(\Phi_c \star
\Phi_c\big) (x) + \textrm{\O}(\hbar^2) \nn \\*
\eea
\end{subequations}
The star products in the previous expressions are to be understood as
follows:
\bea
(\ref{suba}) \!\!\! & \propto & \!\! \Big[\st{\xi}{\eta}\; 
\str{\alpha}{\beta}\; \Phi_c(x+\xi) \int d^4 y \;G(x+\eta +\alpha,y)\,
\frac{\delta}{\delta \Phi_c(y)}\;  \Phi_c(x+\eta +\beta)\Big]_{\{\xi\}
= 0 }  \nn \\
& = & \!\! \Big[ \st{\xi}{\eta}~\str{\alpha}{\beta} ~\Phi_c(x+\xi)
\int d^4 y~ G(x+\eta +\alpha,y) \,\delta(y-x -\eta -\beta)
\Big]_{\{\xi\} = 0 } \nn \\ 
& & \nn \\
& = & \!\! \Big[\st{\xi}{\eta}~\str{\alpha}{\beta}~ \Phi_c(x+\xi)~
G(x+\eta +\alpha, x+\eta +\beta) \Big]_ {\{\xi\} = 0 }\nn \\
& & \nn \\
& = & \!\! \Big[\st{\xi}{\eta}~\str{\alpha}{\beta} ~\Phi_c(x+\xi) \int
\db^4 k~ \tilde{G}(k)~ e^{i k(\alpha -\beta)} \Big]_{\{\xi\} = 0 }
\eea
In the last expression there is no term which is $\eta$ dependent so
from the expansion of the first exponential we only remain with the
first term i.e. 1, while the second exponential will give $
e^{\frac{i}{2} k\te k}$ which is also 1 due to the antisymmetry of
$\te$. So,
\bea
(\ref{suba}) & \propto & \Phi_c(x) \int \db^4 k ~\tilde{G} (k) ~ = ~ 
\Phi_c(x) \; G(0) \nn \\
& = & \Phi_c(x) \; G_0(0) \, .
\eea 
For the second term the calculations go on in the same way and \footnote{Note
that $\phi_c(k)$ are the Fourier modes of $\Phi_c(x)$.}:
\bea
(\ref{subb}) & \propto & \bigg\{ \st{\xi}{\eta} ~ \str{\alpha}{\beta}
\int d^4 y~ G(x+\xi, y) \nn \\ 
& &~~ \times \frac{\delta}{\delta \Phi_c(y)}\,\Big[\Phi_c(x+\eta
+\alpha) \cdot \Phi (x+\eta +\beta)\Big] \bigg\}_{\{\xi\}=0} \nn \\
\nn \\
& = & G(0)\;\Phi_c(x) + \int \db^4 k \db^4 q ~\kstd{k}{q} ~ e^{i qx}
\tilde{G}(k) \;\phi_c(q)\, . \nn \\
\eea
Altogether we can write the one loop contribution to the effective action:
\bea
\label{g1}
\frac{\delta \G_1}{\delta \Phi_c(x)} & = & -\frac{\l}{3}\, G_0(0) \;
\Phi_c(x) - \frac{\l}{6} \int \db^4 k ~ \db^4 q ~ \kstd{k}{q} ~ e^{i q
x} ~ \tilde{G}_0(k) \; \phi_c(q) \, .
\eea

\subsection{Diagrammatics}

To give the diagrammatic expansion of the effective action, first one
should derive the Feynman rules for the \nc{e} $\Phi^4$ theory.
It will not be surprising to say that the usual rules can be 
applied. This is because the free theory is the same as in the
commutative case. So any line will represent a $G_0$ and for a vertex
with $n$ lines coming out one should write the $n$-th order functional
derivative of the classical action. The argument can go further
\bea
\frac{\delta S}{\delta \Phi_c(x_3)}\; G_0(x_1,x_2) & = & \int d^4 y~
d^4 z \;G_0(x_1,y) \;\frac{\delta^3}{\delta \Phi(y) \delta \Phi(z)
\delta \Phi(x_3)} \; G_0(z,x_2) \, ,
\eea
or in diagrammatic and condensed notation:
\bea
\frac{\delta}{\delta \Phi_c^m} \left(\includegraphics[bburx=430,
bbury=370, bbllx=250, bblly=330, width=2.5cm]{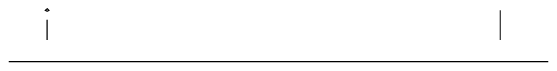}\right) & = &
\includegraphics[bburx=430, bbury=400, bbllx=250, bblly=330,
width=2.8cm]{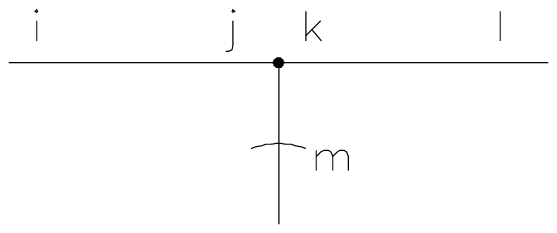}
\eea

Using these conventions, it is easy to verify that we will recover the
Feynman rules we found in the canonical quantization method for \nc{e}
$\Phi^4$ theory. For this, we have to compute the fourth order
functional derivative of the interaction term. This is done in the Appendix
\ref{fd} and the result written in momentum space (\ref{d4sms}),
\bea
\frac{\delta^4 S}{\delta\phi_c(k_1) \; \delta\phi_c(k_2) \; \delta\phi_c(k_3)
  \; \delta\phi_c(k_4)} & \propto & \frac{1}{3} \Big[\c{k_1}{k_2}\;
\c{k_3}{k_4} + 2 \textrm{perm} \Big]\, ,
\eea
is exactly the factor we wrote for the \nc{e} vertex (\ref{factv}).

\subsection{Diagrammatic expansion of the effective action}

We are now ready to write down the diagrammatic expansion of the
effective action for the \nc{e} $\Phi^4$ theory. First we 
show explicitly that up to one loop the diagrammatic expressions
for the commutative and \nc{e} case coincide. To prove this we show
that if we start from 
\bea
\frac{\delta \G_1}{\delta \Phi_c(x)} & = & \frac{1}{2} 
\includegraphics[bburx=470, bbury=400, bbllx=320, bblly=300,
width=1.5cm]{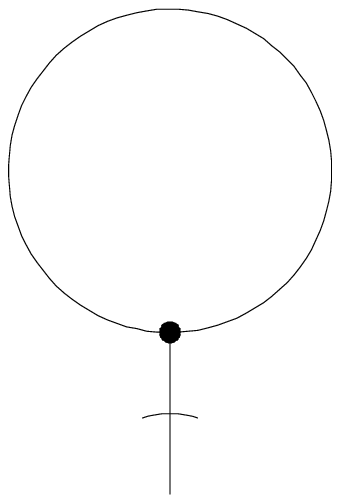} \, ,
\eea
exactly we are going to recover eq. (\ref{g1}). Using the
diagrammatic
rules described in the previous section and the result for the
third order functional derivative of $S$,  (\ref{d3s}) we have:
\bea 
\label{5.14}
\frac{1}{2}
\includegraphics[bburx=470, bbury=400, bbllx=320, bblly=300,
width=1.5cm]{Fig1/circ.eps} & = & \frac{1}{2} \int d^4 y \;d^4 z
~G_0(y,z) \frac{\delta^3 S}{\delta \Phi(y) \delta \Phi(z) \delta
\Phi(x)} \nn \\
\nn \\
& = & -\frac{1}{2}\; \frac{\l}{6}\bigg\{ \st{\xi}{\eta} ~
\str{\alpha}{\beta} \Big[ \; G(x+ \xi,x +\eta + \alpha)\; \Phi(x +
\eta + \beta) \nn \\
& & + G(x + \xi,x +\eta + \beta)\; \Phi(x + \eta + \alpha) 
+ G(x+ \eta + \alpha,x + \xi)\; \Phi(x+ \eta + \beta) \nn \\
& & + G(x+ \eta + \beta,x + \xi) \; \Phi(x+ \eta + \alpha)  
+ G(x+ \eta + \alpha,x+ \eta + \beta) \; \Phi(x + \xi) + \nn \\
& & + G(x+ \eta + \beta,x+ \eta + \alpha) \; \Phi(x + \xi) \,\Big]
\bigg\}_{\{\xi\}=0} \, .
\eea
Written in terms of the Fourier modes (\ref{5.14}) becomes:
\bea
\frac{1}{2}
\includegraphics[bburx=470, bbury=400, bbllx=320, bblly=300,
width=1.5cm]{Fig1/circ.eps} & = & - \frac{1}{2}\; \frac{\l}{6}\;
\bigg\{ \st{\xi}{\eta} ~\str{\alpha}{\beta} ~ \int \db^4 k ~ \db^4 q
~\tilde{G}(k)\; \phi(q) \times \nn \\*
& \times & \Big[ e^{i k(\xi - \eta - \alpha)} \; e^{i q(x
+ \eta + \beta)} + \; e^{i k(\xi - \eta - \beta)} \; e^{i
q(x + \eta + \alpha)} 
+ \; e^{i k(\eta + \alpha -\xi)} \; e^{i q(x + \eta +
\beta)} \nn \\*
& & + \; e^{i k(\eta + \beta -\xi)} \; e^{i q(x + \eta +
\alpha)} + \; e^{i k(\alpha-\beta)} \; e^{i q(x + \xi)} +
e^{i k(\beta - \alpha)} \; e^{i q(x + \xi)}\Big]
\bigg\}_{\{\xi\}=0}\nn \\
& = & - \frac{1}{2}\; \frac{\l}{6} \int \db^4 k ~ \db^4 q
~\tilde{G}(k) \; \phi(q) \; e^{i q x} \Big[ ~e^{- \frac{i}{2}(k\te
q)}~ e^{\frac{i}{2}(k\te q)} + e^{- \frac{i}{2}(k\te q)} ~
e^{\frac{i}{2}(q\te k)} \nn \\*
& & \hspace{4.65cm} + ~e^{\frac{i}{2}(k\te q)} ~ e^{-
\frac{i}{2}(k\te q)} + e^{\frac{i}{2}(k\te q)} ~ e^{-
\frac{i}{2}(q\te k)} + 1 + 1 \Big] \nn \\
& & \nn \\
& = & -\frac{\l}{3}\, G_0(0) \, \Phi_c(x) - \frac{\l}{6} \int \db^4 k
~ \db^4 q ~ \kstd{k}{q} ~ e^{i q x} ~ \tilde{G}_0(k) \; \phi_c(q)
\, . 
\eea
So, up to one loop the diagrammatic expression of the effective action
can be written as: 
\bea
\G[\Phi_c] & = & \bullet~ + ~\left(\frac{\hbar}{i}\right) ~
\frac{1}{2} \includegraphics[bburx=470, bbury=400, bbllx=330,
bblly=330, width=1.7cm]{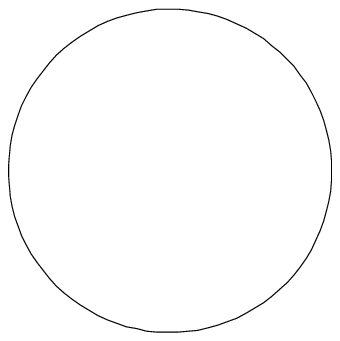} \, .
\eea

\subsection{The effective action at higher orders}

In the previous subsection we explicitly proved that the diagrammatic
expansion up to one loop of the effective action of the \nc{e}
$\Phi^4$ theory is similar to the commutative case. Now we are
going to argue that even at higher orders the effective action should
have the same diagrammatic expansion. This is because in the
calculations we were doing for the commutative theory, the order in
which we were performing the functional integrals was not important,
which is still true for the noncommutative case. Hence we can apply
the same argument to compute the effective action for higher
loops. Therefore, at two loops we have:
\bea
\G[\Phi_c] & = & \bullet~ + ~\left(\frac{\hbar}{i}\right) ~
\frac{1}{2} \includegraphics[bburx=470, bbury=400, bbllx=330,
bblly=330, width=1.7cm]{Fig1/circ0.eps} + ~
~\left(\frac{\hbar}{i}\right)^2 ~
\left[\frac{1}{8} \; \includegraphics[bburx=470, bbury=400, bbllx=310,
bblly=330, width=1.7cm]{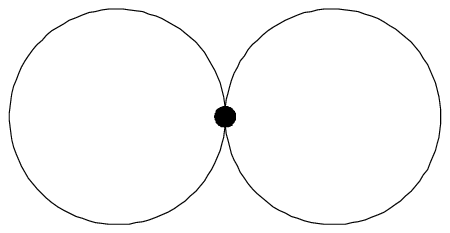} ~ + ~ \frac{1}{12} \; 
\includegraphics[bburx=510, bbury=400, bbllx=310, bblly=375,
width=1.7cm]{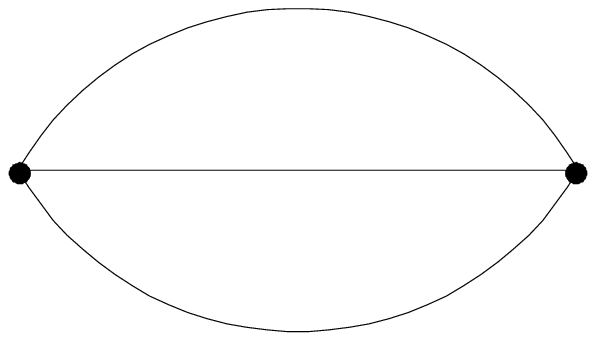} \right] ~ + ~\textrm{\O}(\hbar^3) \,
.\nn \\ 
\eea

\subsection{Planar and nonplanar diagrams}

Here we introduce another way of treating the loop
diagrams \cite{filk} which turns out to be useful in the discussion of
the renormalizability.
Up to now we assumed that the \nc{e} vertex is unique and is
given by eq. (\ref{factv}). 

The alternative point of view is to say
that the order in which the legs appear in the vertices is important
and for every distinct ordering we can associate a phase factor such
that when we sum up over all the possible orderings we recover the
usual factor (\ref{factv}). Let us introduce a notation for a generic
vertex with $N$ legs numbered in an arbitrary order (say clockwise):
\bea
\label{gv}
V(k_1\,\ldots \, k_N) & = & \exp{\left( \frac{i}{2}\sum_{1 \le i
< j \le N} k_i \te k_j \right)} \, .
\eea
Due to the momentum conservation in vertices, this factor is invariant
under cyclic permutations of the legs. 
As far as only this phase factor is concerned one can deduce some rules
so that any graph can be reduced to a generic vertex for which one
can write the phase factor using (\ref{gv}).

\emph{Rule I:} An internal line connecting two different vertices can
be contracted without changing the overall phase factor associated to
the diagram. The important point to keep in mind is \emph{to preserve
the order of the lines}.

\emph{Rule II:} A line starting and ending in the same vertex which is
carrying the momentum $k$ can be removed, but we should also introduce
a phase factor 
\bea
\delta \varphi & = & e^{\pm i k\te p} \, ,
\eea
where $p$ is the algebraic sum of the momenta which are inside the
loop. 

Applying these two rules any graph can be reduced to a generic vertex
in which only the external lines of the original graph enter,
multiplied by some phase factors (these phases appear when we apply the
\emph{Rule II}) which depend on the external, as well as the internal,
momenta of the initial graph. It is obvious that for tree level graphs in
order to find this phase factor we should apply only the \emph{Rule I}, so at
the end the phase factor will depend only on the external momenta. At loop
level however we may find some graphs for which the final phase factor
also depends on the internal momenta. These are called nonplanar
graphs, while those for which the phase factor depends only on the
external momenta are called planar graphs.



\allowdisplaybreaks

\section{Renormalization of \Nc{} $\Phi^4$ Theory}
\label{ren}


In this chapter we study the renormalizability of the
\nc{e} $\Phi^4$ theory up to two loops. We recall from the previous
chapter that in \nc{e} theories we encounter two kinds of diagrams
which are giving the loop corrections: planar and nonplanar
graphs. The planar graphs are the same as the diagrams in the usual
commutative theory, the difference consisting in some numerical and
external momentum dependent  phase factors. For the nonplanar graphs
we find some nontrivial phase factors which depend on the loop momenta
and which can modify dramatically the result of integration. In this
section we are going to show that applying the usual renormalization
procedure for the planar graphs, all the other infinities coming from
the nonplanar diagrams are going to be canceled out, yielding a finite
result, and in this way we prove the renormalizability of the theory
up to two loops.

\paragraph{Notation} We are going to extend the attributes planar and
nonplanar even on mathematical formulae. A term will be called planar
if it does not contain phase factors which depend on the internal
momentum, and nonplanar in the other case.

\subsection{1-loop renormalization of $\G^{(2)}$}

The Euclidean action for the \nc{e} $\Phi ^4$ theory including the
counter-terms can be written as:  
\bea
S & = & \int \db^4 x \left[\frac{1}{2}\partial_\mu\Phi \;
\partial^\mu\Phi + \frac{m^2}{2} \, \Phi^2 +
\frac{\l}{4!}~\Phi\star\Phi\star\Phi\star\Phi \right]\nonumber \\
& + & \int d^4 x \left[\frac{1}{2}(Z_3-1)\,  \partial_\mu\Phi \;
\partial^\mu \Phi + \frac{\delta m^2}{2} \, \Phi^2 +
\frac{\delta\l}{4!}~\Phi\star \Phi\star\Phi\star\Phi \right] \, .
\eea
This leads to the following diagrammatic expansion of $\G^{(2)}$ at one
loop: 
\be
\G ^{(2)} ~ = ~
\includegraphics[bburx=480, bbury=395, bbllx=225, bblly=330,
width=3cm]{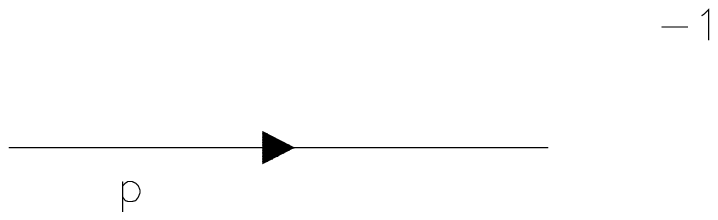}
\quad + \quad
\frac{1}{2}\includegraphics[bburx=440, bbury=460, bbllx=220,
bblly=333, width=3cm]{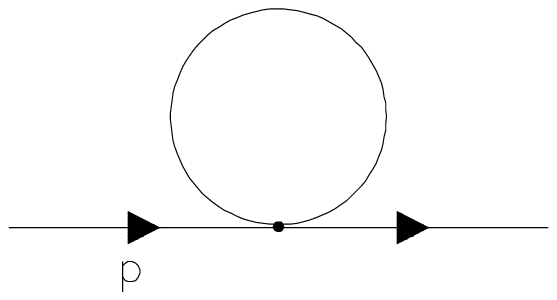}
\quad + \quad
\includegraphics[bburx=480, bbury=395, bbllx=225, bblly=330,
width=3cm]{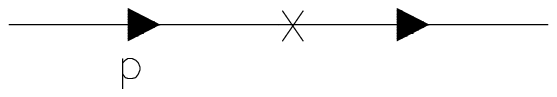} \, .
\end{equation}

The one loop mass correction in the \nc{e} theory has the form:
\bea
\label{mascorr}
\includegraphics[bburx=440, bbury=460, bbllx=220, bblly=350,
width=4cm]{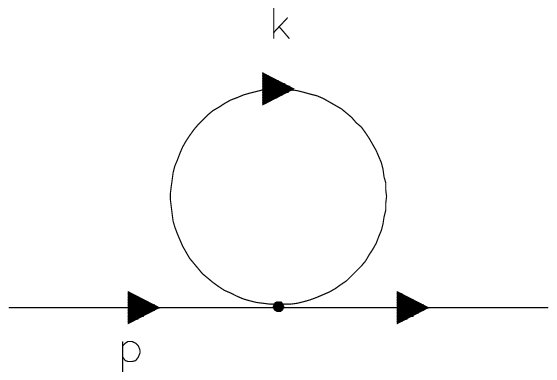} 
& = & -\frac{\l}{3} \int \left[ 2 \left(\cos{\frac{p\te k}{2}}
\right)^2 +1 \right]~ \frac{\db^4 k}{k^2 + m^2} \nn \nonumber\\
& = & -\frac{\l}{3} \int \frac{\cos{p\te k} + 2}{k^2 + m^2}\,
\db^4k \nn \nonumber\\
& = & -\frac{2}{3} \l \int\frac{\db^4 k}{k^2 + m^2}~-~\frac{\l}{3}
\int \frac{\cos{p\te k}}{k^2 + m^2} \, \db^4 k \, .
\eea

Using Schwinger parameterization
\bea
\frac{1}{k^2+m^2} & = & \int_0^\infty d\alpha ~e^{-\alpha(k^2+m^2)} \, ,
\eea
the integral over $k$ becomes just an ordinary Gaussian integral which
can be performed explicitly.
\bea
\includegraphics[bburx=440, bbury=460, bbllx=220, bblly=350,
width=2.5cm]{Fig1/fprop1l.eps} & = & 
-\frac{\l}{3} \int_0^\infty d\alpha \int \db^4k \,\left[ 2\, e^{-
\alpha(k^2 + m^2)} + e^{i k\te p - \alpha(k^2+m^2)} \right] \nn \\
& = & -\frac{\l}{3 \; (2\pi)^4} \int_0^\infty d\alpha
\left(\sqrt{\frac{\pi}{\alpha}}\right)^4 \left[2 \, e^{-\alpha m^2} +
e^{-\alpha m^2 - \frac{p\circ p}{4 \alpha}} \right] \, ,
\eea
in which we have introduced the notation $p\circ k \equiv p \te \te k = p_\mu
\te^{\mu \rho} \te_{\!\!\rho}^{\:\nu} k_\nu \,$. The integral over
$\alpha$ can be regularized by introducing a factor $e^{-\frac{1}{4
\alpha \Lambda^2}}$, where $\Lambda$ plays the role of a cut-off.
\bea
\Bigg[\includegraphics[bburx=440, bbury=460, bbllx=220, bblly=350,
width=2.5cm]{Fig1/fprop1l.eps}\Bigg]_{\textrm{NP}} 
& = & -\frac{\l}{3 \cdot 2^4 \, \pi^2}\int _0^\infty
\frac{d\alpha}{\alpha^2} e^{-\alpha m^2-\frac{1}{4\alpha \lef^2}} =
\nn \\*
& = & -\frac{\l}{12 \pi^2}\,m^2 \, \sqrt{\frac{\lef^2}{m^2}} ~K_1\left(
\frac{m}{\lef} \right) \, ,
\eea
where $\lef^{-2} = p\circ p +\frac{1}{\Lambda^2}$, and $K_1$ is the
modified Bessel function \cite{rijik}. 
It can be seen that in the limit $\Lambda \to \infty \, ,~~ \lef $ becomes
$p\circ p$ so the integral remains finite regulated by the cosine
factor which appears in (\ref{mascorr}).

For the planar part of the diagram we can repeat the calculations with
the change $\lef \to \Lambda$, and then we absorb the regulated
integral in $\delta m ^2$ to make $\G^{(2)}$ finite
\be
\label{dm1}
\delta m_1^2 ~=~ \frac{\l}{3} \int_{\L} \frac{\db^4k}{k^2 + m^2}  \, .
\end{equation}
Here we note that the numeric factor, $\frac{1}{3}$, is different from that of
the commutative case (which is one). Therefore, considering the loop effects,
the $\theta\to 0$ limit is not a smooth limit, and we are not going to recover
the usual commutative theory.

We can write now the quadratic part of $\G^{(2)}$ at one loop:
\bea
\G^{(2)}_1 & = & \int d^4 p \frac{1}{2} \left[p^2 + M^2 +
\frac{\l}{96\, (2\pi)^2} K_1\left(\frac{M^2}{\lef} \right) +
\textrm{\O}(\l^2) \right] \phi(p) \phi(-p) \, .
\eea
Here $M^2$ is the renormalized mass $M^2 = m^2 + \delta m^2_1$.

For small arguments $K_1$ can be expanded in Laurent series
\bea
K_1(z) \xrightarrow{~z\to 0~} \frac{1}{z}  +
\frac{z}{2}\ln{\frac{z}{2}}\, ,
\eea
so that for $\lef \gg 1 $  the quadratic part of the renormalized
effective action is:
\bea
\label{sef2}
\G^{(2)}_{ren}(\Lambda) & = & \int d^4 p\;  \frac{1}{2}~\phi(p)
\; \phi(-p) \times \nn \\*
& \times & \left[p^2 + M^2 +
\frac{\l}{96\, (2\pi)^2 \left(p\circ p + \frac{1}{\Lambda^2} \right)}
- \frac{\l M^2 }{96\, \pi^2} \ln{\left(\frac{1}{M^2\left(p\circ p +
\frac{1}{\Lambda^2} \right)}\right)}
 + \textrm{\O}(\l^2) \right] \nn \\*
\eea

As we see after sending $\Lambda$ to infinity the
$\G^{(2)}$  presents an infrared divergence. Moreover, if we first
consider the zero momentum limit the cut-off effect of the \nc{ity}
cannot be seen any more and the two-point effective action
diverges. This is the interesting UV-IR mixing which appears in the
noncommutative theories. This divergence can be explained assuming
that (\ref{sef2})  is the Wilsonian effective action obtained by
integrating out some other field $\chi$ (see \cite{sei1}). Then the
action which correctly reproduces the factor $\frac{1}{p\circ p}$ in
eq. (\ref{sef2}) is: 
\bea
\G'(\lambda) & = & \G(\Lambda) + \int d^4 x ~\left(\frac{1}{2} \;
\partial \chi \circ \partial \chi + \frac{1}{2} \Lambda^2 (\partial
\circ \partial \chi)^2 + i \sqrt{\frac{\l}{96 \pi^2}}\; \l \chi
\phi \right) \, . \nn \\*
\eea
However the logarithmic term in (\ref{sef2}) is yet to be explained in
some other way \cite{19}.

\subsection{1-loop renormalization of $\G^{(4)}$}

The one loop diagrammatic expansion of $\G^{(4)}$ is:
\be
\label{g4}
\G^{(4)} ~=~ 
\includegraphics[bburx=400, bbury=350, bbllx=225, bblly=350,
width=2cm]{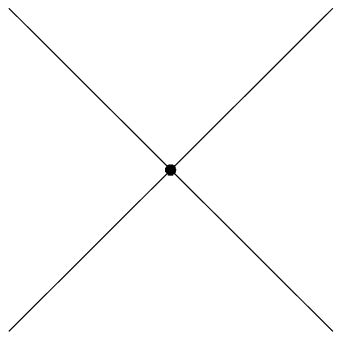}
\quad + 
\includegraphics[bburx=400, bbury=350, bbllx=225, bblly=350,
width=2cm]{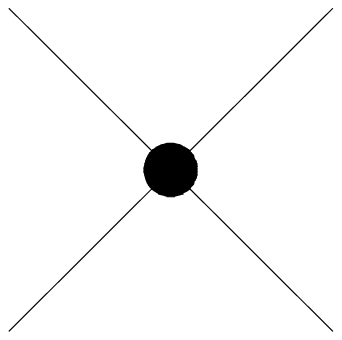}
\quad + 
\frac{1}{2}\left(\includegraphics[bburx=590, bbury=470, bbllx=230,
bblly=370, width=3cm]{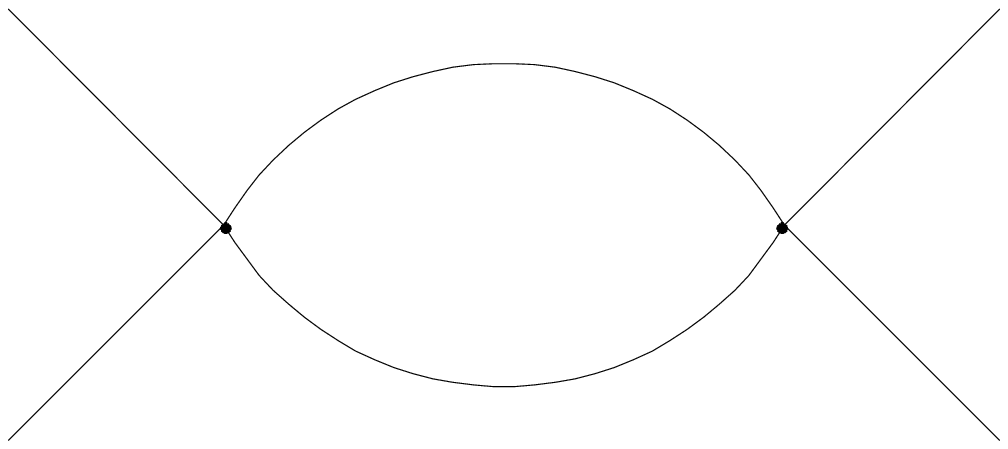} ~+~ 2~~\textrm{permutations}
\right) \, .
\end{equation}


The nonplanar part of the loop graph give rise to an integral of the form
\footnote{The exact phase factors which enter the integral can be found in
Appendix \ref{pf}}:
\bea
\label{npint}
I_{np} & \equiv & \int \db^4k
\frac{e^{i p\te k}}{(k^2+m^2)\big((p_1+p_2-k)^2+m^2\big)} \, ,
\eea
where $p$ is the total momentum which is crossing the loop. As we
showed for $\G^{(2)}$, the exponential factor acts as a regulator and
the integral remains finite. Moreover, the integral can be evaluated
analytically using first Feynman parameterization in order to write the
denominator as a square and then using Schwinger parameterization in order to
compute the integral over the internal momentum and the final result is:
\bea
I_{np} & = & \frac{1}{8 \pi^2} \int_0^1 dx~ e^{i \big(p\te
(p_1+p_2) \big)\,x} ~ K_0\left(\sqrt{\Big[(p_1+p_2)^2\,x\;(1-x) + m^2
\Big] \; p\circ p} \right) \, .
\eea
There is still the peculiar IR
divergence, but in the following we are going to study only the UV
behavior of the theory. So the infinity comes only from the planar
part and it is of the form:
\be
\frac{\l^2}{9}\int \frac{\db^4k}{(k^2+m^2)\big((p_1+p_2-k)^2+m^2\big)}
\cos{p_1\te p_2} \cos{p_3\te p_4} +~ 2~\textrm{permutations} \, .
\end{equation}

Renormalization of $\G^{(4)}$ requires to absorb the infinity in the
corresponding counter-term in eq. (\ref{g4}), i.e.
\bdm
\left. \left.
\includegraphics[bburx=400, bbury=410, bbllx=225, bblly=350,
width=2.5cm]{Fig1/vertct.eps}\right|_{p=0}
~+~\frac{3}{2}~\textrm{Planar} \left[\includegraphics[bburx=590,
bbury=480, bbllx=230, bblly=370, width=2.5cm]{Fig1/fvert1l.eps} \right]
\right|_{p=0} ~= ~ 0 
\edm

\bea
\Rightarrow ~~ \delta\l_1 & + & \frac{3}{2}\cdot \frac{2\l^2}{9}
\int_\L \frac{\db^4 k}{(k^2 + m^2)^2}  = ~ 0 \nn \\
\Rightarrow ~~ \delta\l_1 & = & -~\frac{\l^2}{3}\int_\L \frac{\db^4
k}{(k^2 + m^2)^2} \, .
\eea
We again note that the difference in the numeric factor of $\frac{-1}{3}$ in
the above equation compared to the commutative one which is $-\frac{3}{2}$.

At this point using the low external momenta limit we can 
explicitly write the renormalized effective action at one loop.
For small arguments, the modified Bessel function behaves like:
\bea 
K_0(x)~ \xrightarrow{~x\to 0~} -\ln{\frac{x}{2}} + \textrm{finite} \, ,
\eea
so we can write:
\bea 
\label{inp}
I_{np} & \sim & \frac{1}{16\;\pi^2}~\ln{\frac{4}{m^2 \; p\circ p}} \, .
\eea

Now,
\bea
\left[\frac{1}{2}\includegraphics[bburx=590, bbury=480, bbllx=230,
bblly=370, width=3cm]{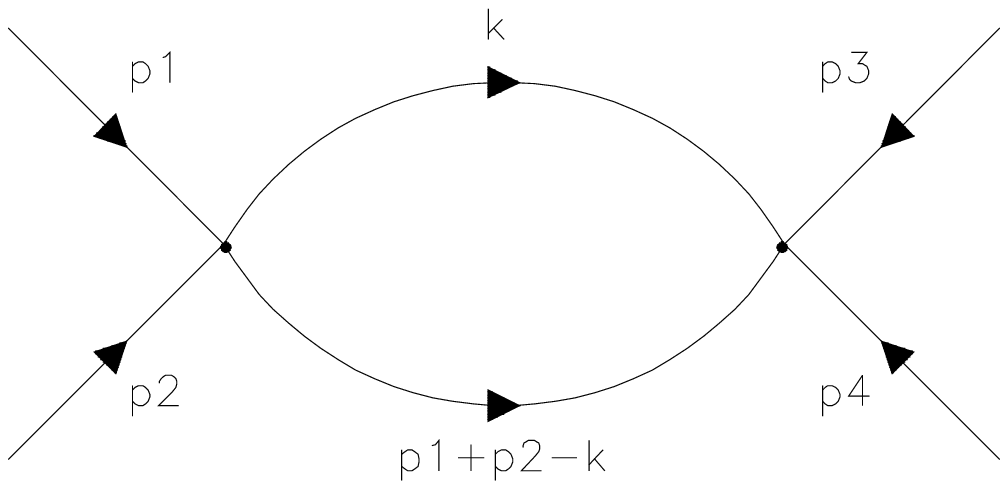} \right] _{\substack{
\textrm{NP} \\ p_i \to 0}} & = & \frac{\l^2}{9} \int \db^4k
\frac{\mathcal{P}(p_1 \ldots p4,k, \te)}{(k^2+m^2) \big((p_1+p_2-k)^2 +
m^2\big)}\; ,
\eea
where $\mathcal{P}$ is given in (\ref{cf1l}). 
In the limit which we are considering one can neglect the cosine
factors which depend only on the external momenta. We should also notice that
$I_{NP}$ does not depend on the sign in the exponential and so taking into
account the ``2 permutations'' from eq. (\ref{g4}) we can write:
\bea
\left[\frac{1}{2}\includegraphics[bburx=590, bbury=480, bbllx=230,
bblly=370, width=3cm]{Fig1/vert1l.eps} + 2\textrm{perm} \right] _{
\substack{\textrm{NP} \\ p_i \to 0}} & = & \frac{\l^2}{9} \int \db^4k
\frac{1}{(k^2+m^2)\big((p_1+p_2-k)^2+m^2\big)}\; \nn \\*
& \times & \left[\sum_{i=2}^{4}  e^{i k\te (p_1 + p_i)} ~ + ~
\frac{3}{2} \sum_{i=1}^{4}  e^{i k\te p_i} ~ + ~ \frac{2}{4}
\sum_{i=2}^{4}  e^{i k\te (p_1 + p_i)} \right] \, . \nn \\*   
\eea
Now using (\ref{inp}) we can write the renormalized four point
function at one loop: 
\bea
\G^{(4)}_{ren}(p_1\ldots p_4) & = & \l - \frac{\l^2}{96\, \pi^2} \bigg\{
\ln{\frac{1}{m^2 \; p_1\circ
p_1}} + \ln{\frac{1}{m^2 \; p_2\circ p_2}} + \ln{\frac{1}{m^2 \;
p_3\circ p_3}}  \nn \\
& & \hspace{1.9cm}+ \ln{\frac{1}{m^2 \; p_4\circ p_4}} +
\ln{\frac{1}{m^2 \; (p_1+ p_2)\circ (p_1+p_2)}} \nn \\
& & \hspace{1.9cm}+ \ln{\frac{1}{m^2 \; (p_1+ p_3)\circ (p_1+ p_3)}} +
\ln{\frac{1}{m^2 \; (p_1+ p_4)\circ (p_1+ p_4)}} \bigg\}\,. \nn \\ 
\eea

\subsection{$\G^{(2)}$ at two loops}
\label{sg22}

After the one loop calculation, we can proceed to that of two
loops. First we start with two point function:
\bea
\label{g22}
\G^{(2)} & = & 
\includegraphics[bburx=480, bbury=395, bbllx=225, bblly=330,
width=3.5cm]{Fig1/prop.eps} +
\includegraphics[bburx=480, bbury=395, bbllx=225, bblly=330,
width=3.5cm]{Fig1/propct.eps} +
\includegraphics[bburx=480, bbury=395, bbllx=225, bblly=330,
width=3.5cm]{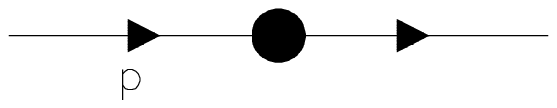}~ + 
\nn \\*
& & \hspace{1.1cm} (a) \hspace{3.6cm} (b) \hspace{3.5cm}
(c) \nn \\ 
& + & \frac{1}{2}~\includegraphics[bburx=440, bbury=460, bbllx=225,
bblly=350, width=3.5cm]{Fig1/fprop1l.eps} 
+\quad \frac{1}{2}
\includegraphics[bburx=440, bbury=395, bbllx=225, bblly=350,
width=3.5cm]{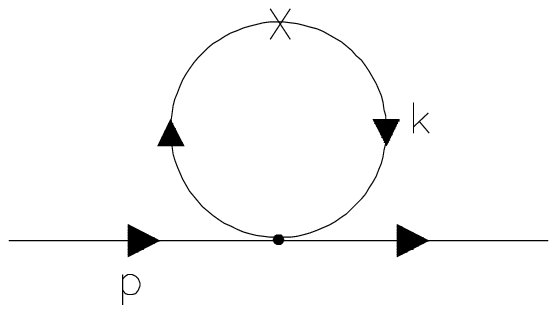} + \quad \frac{1}{4}
\includegraphics[bburx=400, bbury=460, bbllx=220, bblly=350,
width=3.1cm]{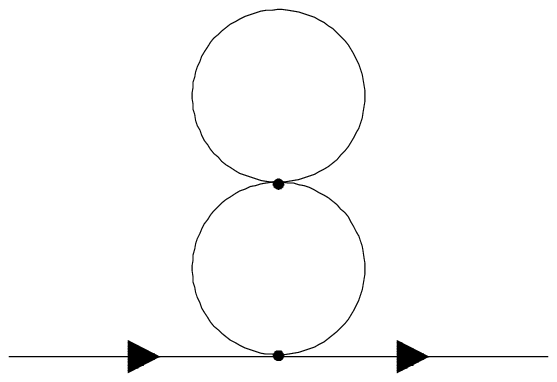} \nn \\*
& & \hspace{2cm} (d) \hspace{4.2cm} (e) \hspace{4.5cm} (f) \nn \\
& + & \frac{1}{2}~ 
\includegraphics[bburx=480, bbury=450, bbllx=225, bblly=350,
width=4cm]{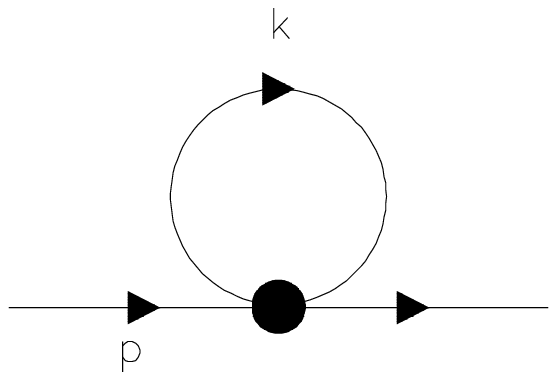}
+ \quad \frac{1}{6}~
\includegraphics[bburx=580, bbury=460, bbllx=250, bblly=380,
width=4cm]{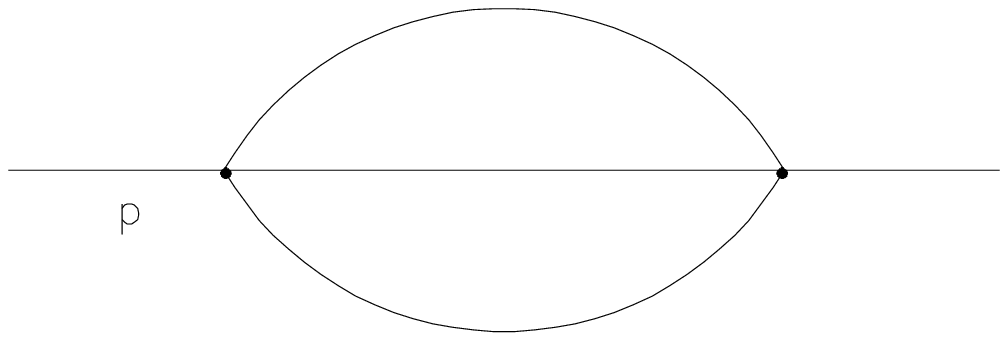} \, . \nn \\*
& &\hspace{2cm} (g) \hspace{6cm} (h) \nn \\
\eea

In the usual commutative $\Phi^4$ theory at two loops the UV divergent parts of 
terms (e) and (f) cancel out. However, since in the 1-loop mass correction only the
planar part have been taken into account (see eq. (\ref{dm1})),
in the \nc{e} theory we are left with another term.

\bea
\frac{1}{2}~ \includegraphics[bburx=440, bbury=395, bbllx=225,
bblly=350, width=3.5cm]{Fig1/propct4.eps} & =&
\frac{\l^2}{18}\int \db^4 k \; \db^4 q ~\frac{\cos{k\te p} +
2}{(q^2+m^2) (k^2+m^2)^2} \nn \\
\frac{1}{4}~\includegraphics[bburx=440, bbury=460, bbllx=220, 
bblly=350,width=3.5cm]{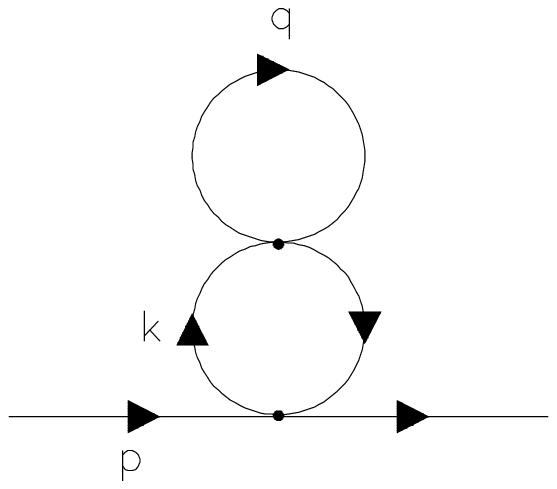} & = &
-\frac{\l^2}{36}\int \db^4 k \; \db^4 q ~\frac{(\cos{k\te p
}+2)(\cos{k\te q} + 2)} {(q^2+m^2) (k^2+m^2)^2} \nn \\
\frac{1}{2}\includegraphics[bburx=440, bbury=395, bbllx=225,
bblly=350, width=3.2cm]{Fig1/propct4.eps} +
\frac{1}{4}\includegraphics[bburx=440, bbury=460, bbllx=225, 
bblly=350,width=3.2cm]{Fig1/prop2l.eps} & = &
-\frac{\l^2}{36}\int \db^4 k \; \db^4 q ~ \frac{\cos{k\te q}(\cos{k\te
p}+2)} {(q^2+m^2) (k^2+m^2)^2}\, . \nn \\ 
\eea

Using Schwinger parameterization, the integral over $q$ can be done
explicitly and we remain with:
\bea
\frac{1}{2}~ \includegraphics[bburx=440, bbury=395, bbllx=225,
bblly=350, width=2.5cm]{Fig1/propct4.eps} +
\frac{1}{4}~\includegraphics[bburx=440, bbury=460, bbllx=220, 
bblly=350,width=2.5cm]{Fig1/fprop2l.eps} & = &
-\frac{\l^2}{36}\int \db^4 k \int_0^\infty \frac{d \alpha}{(2\pi)^4}
\left(\sqrt{\frac{\pi}{\alpha}}\right)^4 e^{-\alpha m^2-\frac {1}{4
\alpha \lef}} \nn \\
\nn \\
& & \times ~ \frac{\cos{p\te k}+2}{(k^2+m^2)^2} \, , 
\eea
where $\lef$ is given by $ \lef^{-2} = k\circ k
+\frac{1}{\Lambda^2}$. The integral over $\alpha$ can be exactly computed,
yielding a modified Bessel function $K_1$ and the final result
is: 
\bea
\frac{1}{2}~ \includegraphics[bburx=440, bbury=395, bbllx=225,
bblly=350, width=3cm]{Fig1/propct4.eps} +
\frac{1}{4}~\includegraphics[bburx=440, bbury=460, bbllx=220, 
bblly=350,width=3cm]{Fig1/fprop2l.eps} ~= \nn \\
\nn \\
\label{dnp1}
= ~-\;\frac{\l^2 m^2}{36}\frac{1}{2^4 \pi^2}\int \db^4k ~ \frac{4}{
\sqrt{m^2(k\circ k +\frac{1}{\Lambda^2})}} K_1 \left( 
\sqrt{\frac{m^2}{\lef^2}} \right) \frac{cos{p\te k} + 2}{(k^2 + 
m^2)^2} \nn \\
\eea
Simple power counting tells us that the integration over the large
values of $k$ is finite provided $K_1$ does not diverge at infinity.
In fact $K_1$ decays exponentially at infinity, so the convergence is
guaranteed. On the other hand the integration over small values of
$k$ can be controlled if we do not let $\L$ to go to infinity. 
It is worth noting that in the massless limit ($m\to 0$) all these arguments again
holds. In fact in the $m\to 0$ limit all the mass dependence is removed from
(\ref{dnp1}).  
Under these assumptions we can write:
\be 
\frac{1}{2}~
\includegraphics[bburx=480, bbury=395, bbllx=225, bblly=350,
width=3.5cm]{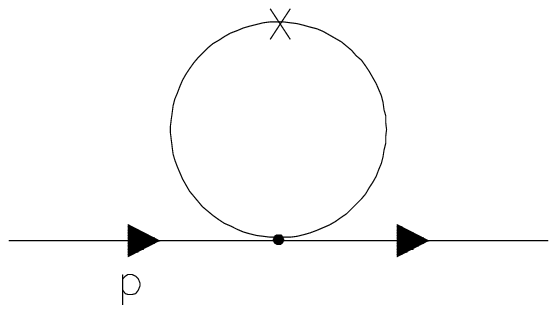} 
\quad + \quad \frac{1}{4}
\includegraphics[bburx=440, bbury=460, bbllx=220, bblly=350,
width=3cm]{Fig1/fprop2l.eps} ~ =~ \textrm{finite}
\end{equation}

In the usual $\Phi^4$ theory,
for the remaining loop diagrams in eq. (\ref{g22}), the momentum
independent infinities are absorbed in $\delta m^2$, while the
momentum dependent ones are absorbed in the wave function
renormalization. 

In \nc{e} $\Phi^4$ theory, the momentum dependent factors which appear
in the vertices for (h) in eq. (\ref{g22}) are:
\be
\frac{1}{9}\left[\c{p}{k}\c{q}{(p-k)} + \c{p}{q}\c{k}{(p-q)} +
\c{k}{q}\c{p}{(k+q)}\right]^2 
\end{equation}

Expanding both the square and the cosine factors we get:
\bea
& & \frac{1}{6}~
\includegraphics[bburx=570, bbury=460, bbllx=260, bblly=380,
width=3.5cm]{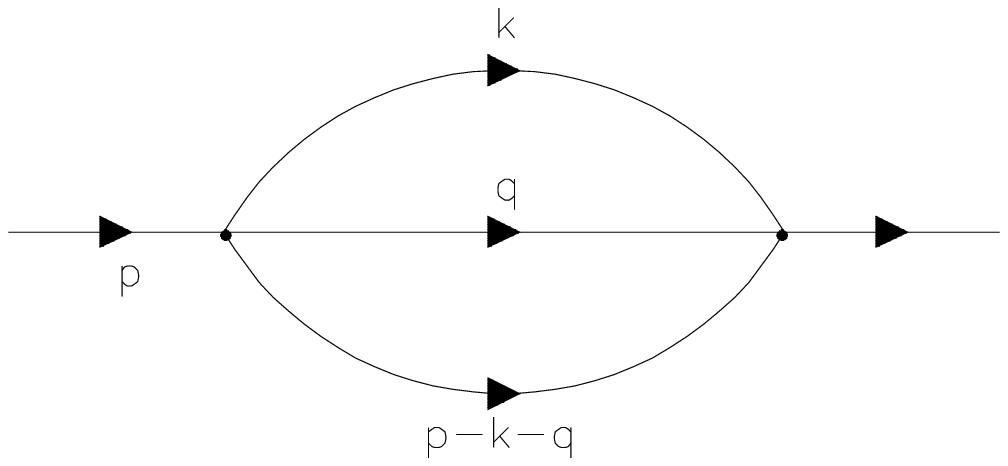} ~ = ~
 -  \frac{\l^2}{36}\int\frac{\db^4 k ~\db^4 q}{(k^2+m^2)(q^2+m^2) 
((p-k-q)^2+m^2)}~\bigg\{1+ \nn \\
\nn \\
& & \quad +\frac{2}{3}\Big[\cos{p \te k} + \cos{p \te q} + \cos{p \te
(k+q)}\Big] + \frac{2}{3}\Big[\cos{k \te (p-q)} + \cos{q \te (p-k)}
+ \cos{k \te q}\Big] \nn \\
& & \quad + \frac{1}{3}\Big[\cos{\big(p \te k +q\te (p-k)\big)}
+ \cos{\big(p \te k - q\te (p-k) \big)} + \cos{\big(p \te k -
q\te(p+k)\big)}\Big]\bigg\} = \nn \\ 
\nn \\
& = & - \frac{\l^2}{36}\int\frac{\db^4 k \db^4 q}{(k^2+m^2)(q^2+m^2) 
((p-k-q)^2+m^2)} ~ \Big[1 + 2 \cos{p\te k} + 2 \cos{k\te q} + \nn \\
& & \hspace{8.55cm} + \; \cos{\big(p\te k + q\te (p-k)\big)}\;\Big] \,
.\nn \\ 
\eea

The contribution of the counter-term (g) in eq.(\ref{g22})  is:

\bea
\quad \frac{1}{2}~ 
\includegraphics[bburx=430, bbury=395, bbllx=225, bblly=350,
width=2.8cm]{Fig1/propct3.eps} & = & - \frac{\l^2}{2}\left(-\frac{1}{3}
\right) \int \frac{\db^4 k}{k^2+m^2} \frac{2+\cos{p\te k}}{3} \int 
\frac{\db^4 q}{(q^2+m^2)^2} = \nn \\
\quad & \quad & \nn \\
& = & \frac{\l^2}{9} \int \frac{\db^4k~ \db^4 q}{(k^2+m^2)(q^2+m^2)^2} + 
\frac{\l^2}{18} \int \db^4k~ \db^4 q\frac{\cos{p\te k}}{(k^2+m^2)
(q^2+m^2)^2} \, . \nn \\
\eea

For the planar diagrams we should follow the same renormalization
procedure as in the commutative case. 
More precisely there are new divergences arising from the planar parts of the sum of
the above two graphs which should be absorbed in the relevant two loop counter-term. 
We again note that this counter-term is not $\theta$ dependent.
We are then left with the nonplanar part which can be written as:

\bea
&  & \quad \left[\frac{1}{2}~ 
\includegraphics[bburx=480, bbury=450, bbllx=225, bblly=350,
width=3cm]{Fig1/propct3.eps}
+ \quad \frac{1}{6}~
\includegraphics[bburx=580, bbury=460, bbllx=250, bblly=380,
width=4cm]{Fig1/prop2l2.eps} \right]_{\textrm{NP}}~= \nn \\*
& & \nn \\*
& = & \frac{\l^2}{18} \int \db^4k~ \db^4 q\frac{\cos{p\te k}}{(k^2+m^2)
(q^2+m^2)} \left[\frac{1}{q^2+m^2} - \frac{1}{(p-k-q)^2+m^2}\right] - 
\nn \\*
& & \nn \\*
& - & \frac{\l^2}{36}\int \frac{\db^4 k \db^4 q}{(k^2+m^2)(q^2+m^2)
((p-k-q)^2+m^2)} \Big[2\cos{k\te q} +\cos{\left(p\te k + q\te(p-k) 
\right)}\Big] \, . \nn \\ 
\eea
For the first term, the integral over $q$ yields a finite result (this
can be seen by simple power counting), while the term $\cos{p\te k}$
acts as a regulator for the integral over $ k $. In the second term
when we take $\cos{k\te q}$ from the square bracket and integrate over
$q$, we 
get a modified Bessel function ($K(\sqrt{k\circ k})$) which
exponentially decay at infinity and takes care of the integration over
large values of $q$. When we take $\cos{(p\te k + q\te(p-k))}$ by a
change of variables ($k'=p-k$ and $q'=p-q$) we can put the integral
in the form: 
\bdm 
\int \frac{\db^4k~ \db^4 q}{((p-k)^2+m^2)((p-q)^2+m^2)((p-k-q)^2 +m^2)}
\cos{k\te q}
\edm
for which, in the UV-limit  we can apply the same argument as before.

At this point we have proved that renormalizing the planar part of the
diagrams appearing in the \nc{e} version of the $\Phi^4$ theory, as in
the usual case we can make $\G^{(2)}$ finite without renormalizing any 
other parameter, in particular, $\te$.

\subsection{$\G^{(4)}$ at two loops}

\bea
\label{g42}
\G^{(4)} & = &
\includegraphics[bburx=400, bbury=350, bbllx=225, bblly=350,
width=2.5cm]{Fig1/fvert.eps} + 
\frac{1}{2}~\left(\includegraphics[bburx=590, bbury=480, bbllx=230,
bblly=370, width=3cm]{Fig1/fvert1l.eps} ~+~ 2~~\textrm{perm}
\right) + 
\includegraphics[bburx=400, bbury=350, bbllx=225, bblly=350,
width=2.5cm]{Fig1/vertct.eps} + \nn \\*
& & \hspace{.8cm} (A) \hspace{3.7cm} (B) \hspace{4.8cm} (C) 
\nn \\ 
& + & \left(\includegraphics[bburx=590, bbury=480, bbllx=230, bblly=370,
width=2.5cm]{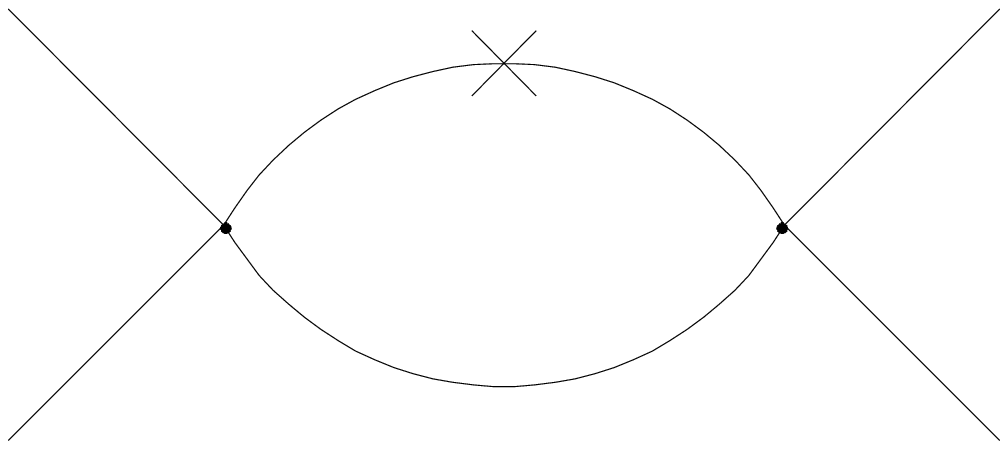} + 2~\textrm{perm} \right)
~ +~ 
\frac{1}{2}\left(\includegraphics[bburx=590, bbury=480, bbllx=230,
bblly=370, width=3cm]{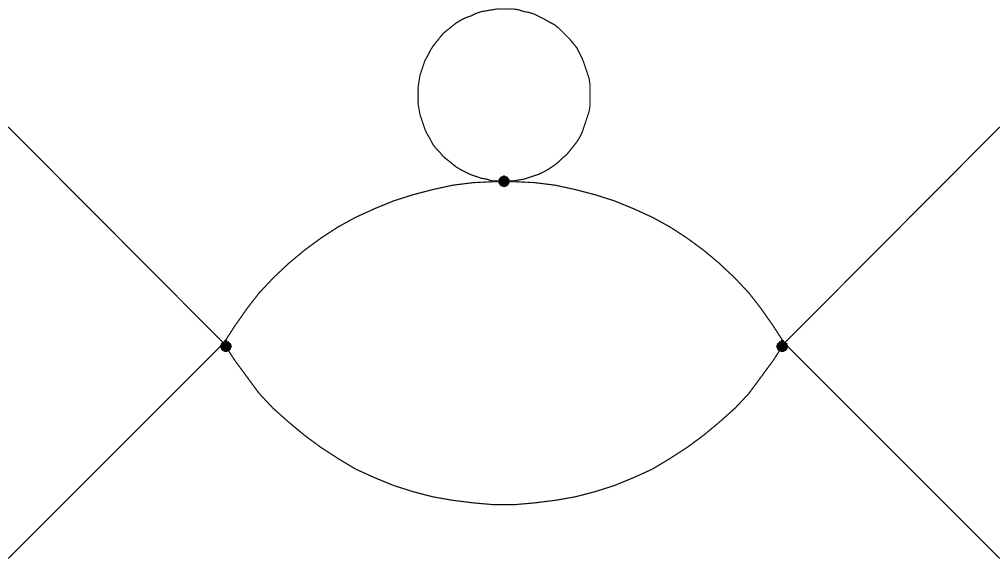} + 2~\textrm{perm}
\right) + \nn \\*
& & \hspace{1.3cm}(D) \hspace{5.7cm} (E)  \nn \\
& + & \frac{1}{2}\left(\includegraphics[bburx=590, bbury=480,
bbllx=230, bblly=370, width=3cm]{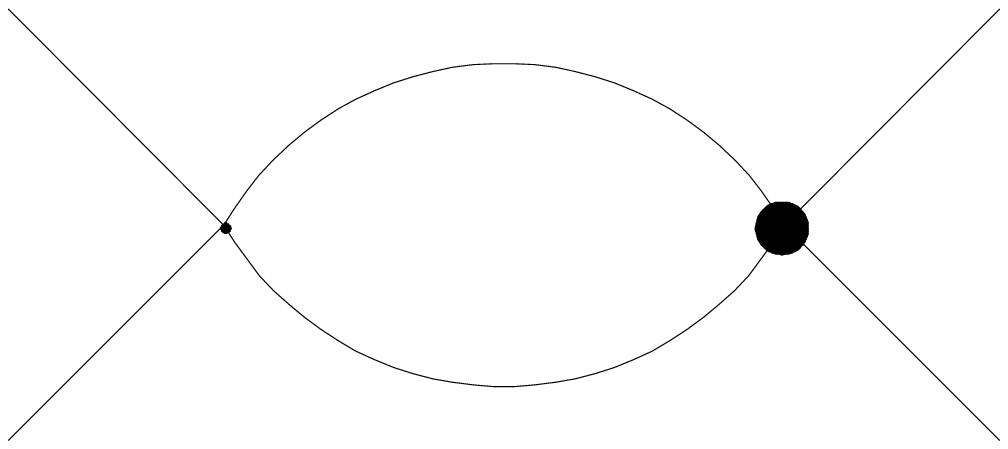} + 5~\textrm{perm}
\right)  + 
\frac{1}{4}\left(\includegraphics[bburx=500, bbury=410, bbllx=270, 
bblly=335, width=3.3cm]{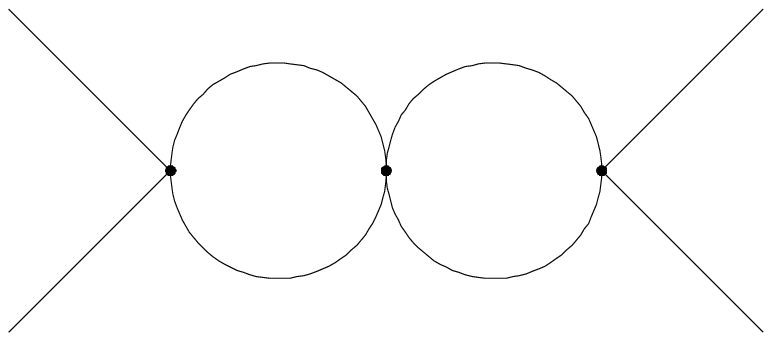} +
2~\textrm{perm} \right) + \nn \\*
& & \hspace{1.7cm}(F) \hspace{6cm} (G)  \nn \\
& + & \frac{1}{4} \left(\includegraphics[bburx=510, bbury=430, bbllx=200,
bblly=340, width=3.3cm]{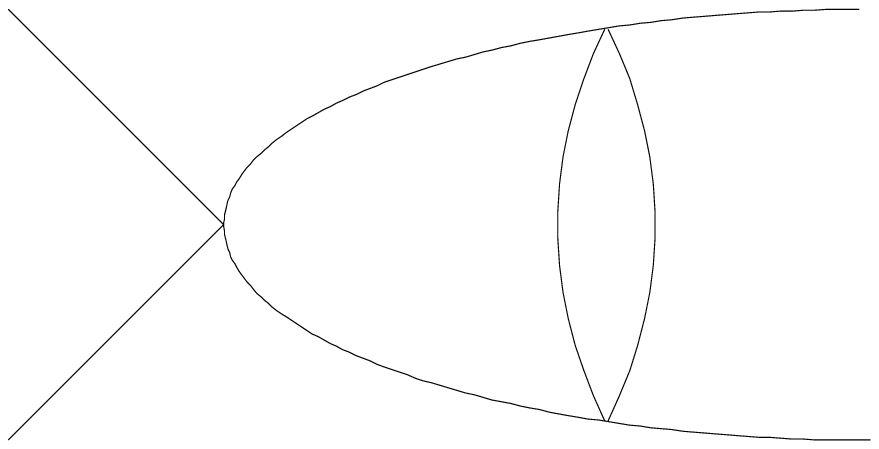} +
11~\textrm{perm} \right) \nn \\*
& & \hspace{1.5cm} (H) \nn \\
\eea

This formula requires some comments. The last term (or the fish
diagram) appears twelve times according to the number of permutations 
of the external momenta which give different contributions. 
In the commutative case however we can see only six independent
permutations. This difference comes from the fact that in the \nc{e}
theory there are momentum dependent phase factors which appear in
vertices, and these factors allow us to distinguish between
the last two legs of the fish diagram. Since all we are doing is to
take the fourth order functional derivative of the effective action,
and the order in which we perform the derivatives has no importance,
in the end when we sum up the diagrams coming from all the
permutations we should find that the result is invariant under
arbitrary relabeling of external momenta. This is the reason why we
need 11 permutations in the last term of eq. (\ref{g42}). However,
since not all the terms in the fish diagram break explicitly the
symmetry between the last legs, we shall consider for simplicity only
5 permutations, but in the end we should remember to symmetrize over
the last two momenta.

In the commutative case divergences coming from terms (D) and (E) of eq.
(\ref{g42}) cancel each other. In the \nc{e} case, because of changes in the numeric
factors of counter-terms, should be checked again.
Using the notation we introduced in eq. (\ref{v1l}) for 
the cosine factors appearing in the 1-loop vertex correction, and also
the definition of $\delta m_1^2$ from eq. (\ref{dm1}), we can write:

\bea
& & \left(\includegraphics[bburx=550, bbury=480, bbllx=230, bblly=370,
width=3.5cm]{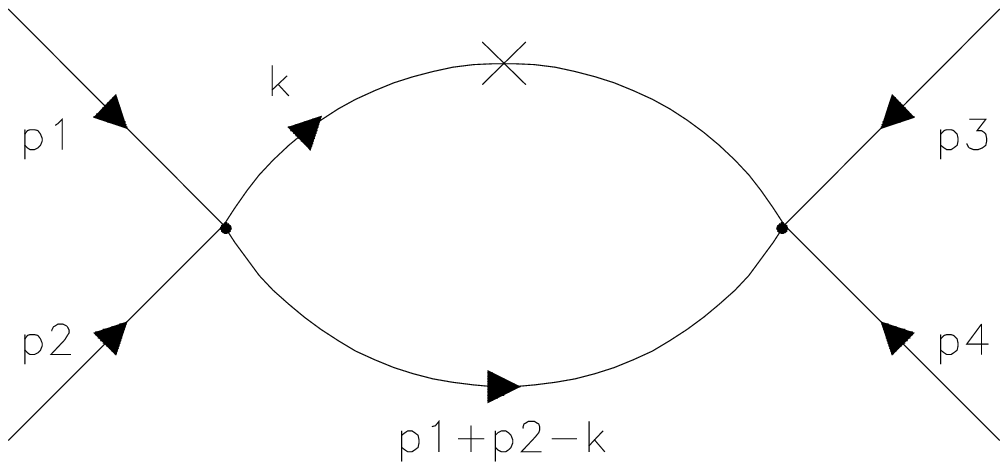} ~+~ 2~~\textrm{perm} \right)
\quad + \quad
\frac{1}{2}\left(\includegraphics[bburx=570, bbury=480, bbllx=230,
bblly=370, width=3.5cm]{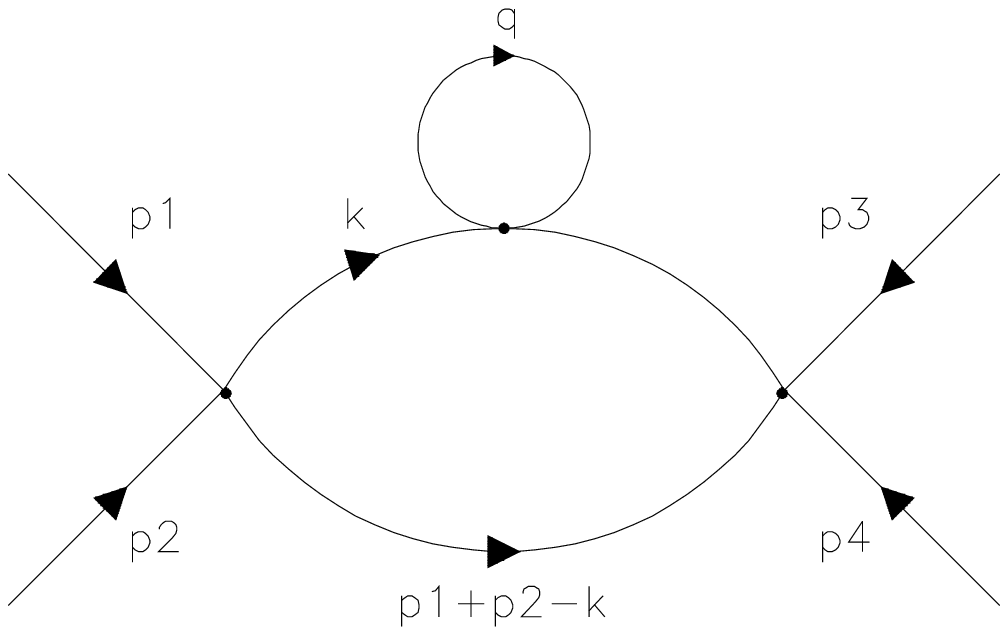} +~ 2~~\textrm{perm}
\right) ~= \nn \\
& & \nn \\
& = & 2\; \frac{\l^2}{9}\int \db^4 k ~\db^4 q\frac{\mathcal{P}(k,p,\te)}
{(q^2+m^2)(k^2+ m^2)^2((p-k)^2 + m^2)} \cdot \frac{\l}{3} - \nn \\
& \quad & \quad \nn \\
& &\hspace{1cm} - 
\frac{\l^2}{9}\int \db^4 k ~\db^4 q
\frac{\mathcal{P}(k,p,\te)} {(q^2+m^2)(k^2+ m^2)^2((p-k)^2 + m^2)} 
\cdot \frac{\l \, \left(cos{k\te q} + 2\right)}{3} ~ =\nn \\
& \quad & \quad \nn \\
& & \hspace{.8cm}=~- 
\frac{\l^3}{27}\int \db^4 k ~\db^4q
\frac{\cos{k\te q}}{q^2+m^2}  
\cdot \frac{\mathcal{P}(k,p,\te)}{(k^2+m^2)^2((p-k)^2+m^2)} \, .
\eea

The $q$ integral is regulated by $\cos{k\te q}$, while the integral
over $k$ is convergent right from the beginning. This means that even
though the sum of these diagrams is nonzero at least it is finite, and
this is what we are interested in.

The planar part of the diagrams in (\ref{g42}) does not come with
anything new, except for some numeric and phase factors which depend
only on the external momenta. Nevertheless in order to apply the usual
renormalization procedure we should check explicitly that the external
momentum dependent factor is the same for all the diagrams which
appear in the expansion of $\G^{(4)}$ and this should be exactly the
additional phase factor for a \nc{e} vertex, i.e. 
\be
\c{p_1}{p_2}\c{p_3}{p_4}+ \c{p_1}{p_3}\c{p_2}{p_4} + 
\c{p_1}{p_4}\c{p_2}{p_3} \, .
\end{equation}
The whole calculation with the explicit cosine expansion is given in the
Appendix \ref{pf}.

Due to the internal momentum phase factors, the nonplanar diagrams are
less divergent than the corresponding planar ones. However
divergences may still appear whenever the cosine factors do not
contain any of the loop momenta. In the following we are going to show that these
infinities coming from the nonplanar graphs are going to cancel so that in
order to obtain an overall finite result it is enough to renormalize the
planar diagrams.

The nonplanar divergent parts of the diagrams (G), (H) in (\ref{g42}) are
obtained by simply taking the terms which we denoted in the Appendix \ref{pf}
as "nonplanar k/q independent terms". Keeping track of all the numerical
factors we find for the diagram (G):
\bea
& & \left. \hspace{-1cm}
\frac{1}{4}\includegraphics[bburx=500, bbury=400, bbllx=250,
bblly=330, width=3.5cm]{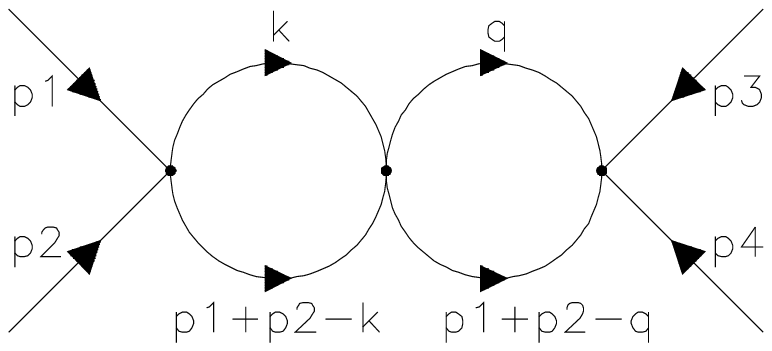} \right|_{\textrm{div.
NP}} ~ =\nn \\
& = & \frac{\l^3}{27} \int \frac{\db^4 k ~\db^4 q}{(k^2 + m^2)
((p_1+p_2 -k)^2 + m^2)(q^2 + m^2)((p_1+p_2-q)^2 +m^2)} \nn \\
& \times & \Bigg\{\frac{1}{2} \c{p_1}{p_2}\c{p_3}{p_4} \bigg[\cos{k\te
(p_1+p_2)} + \cos{q\te (p_1+p_2)}\bigg] + \nn \\*
& & +\frac{1}{4} \c{p_1}{p_2}\bigg[\cos{\left(\frac{p_3\te p_4}{2} -
q\te p_3 \right)}
+ \cos{\left(\frac{p_3\te p_4}{2} + q\te p_4\right)}\bigg]+ \nn \\*
& & +\frac{1}{4} \c{p_3}{p_4}\bigg[\cos{\left(\frac{p_1\te p_2}{2} +
k\te p_1 \right)} + \cos{\left(\frac{p_1\te p_2}{2} - k\te
p_2\right)}\bigg] \Bigg\}  
\eea
Since the propagators corresponding to this diagram are
$q~\longleftrightarrow~k$ symmetric, by a change of variables $q$ can
be replaced by $k$ inside the cosine factors.
\bea
\label{dvnp1}
& & \left. \hspace{-1cm}
\frac{1}{4}\includegraphics[bburx=500, bbury=400, bbllx=250,
bblly=330, width=3.5cm]{Fig1/vert2l2.eps} \right|_{\textrm{div.
NP}} ~= \nn \\
& = & \frac{\l^3}{27} \int \frac{\db^4 k ~\db^4 q}{(k^2 + m^2)
((p_1+p_2 -k)^2 + m^2)(q^2 + m^2)((p_1+p_2-q)^2 +m^2)} \times \nn \\
& \times & \Bigg\{ \c{p_1}{p_2} \c{p_3}{p_4} \cos{k\te (p_1+p_2)} 
+\frac{1}{4} \c{p_1}{p_2}\Bigg[\cos{\left(\frac{p_3\te p_4}{2} - k\te
p_3 \right)} \nn \\*
& + & \cos{\left(\frac{p_3\te p_4}{2} + k\te p_4\right)}
\Bigg]~+
~\frac{1}{4} \c{p_3}{p_4}\bigg[\cos{\left(\frac{p_1\te
p_2}{2} + k\te p_1 \right)} + \cos{\left(\frac{p_1\te p_2}{2} - k\te
p_2\right)} \bigg] \Bigg\}~ \nn \\*
\eea

Now we should do the same with the fish diagram. However here we should notice
that the propagators in diagram (H) \footnote{See the picture below for the
right assignment of momenta} come with a $k^6$, so the integration
over $k$ is already UV finite despite of the fact that there is no regulator
on the $k$ integral coming from the noncommutativity. This means that the
nonplanar k-independent terms will give a finite result because they are
regulators for the q-integral. So the only nonplanar terms which are
contributing to the divergences of the fish diagram come only from the so
called q-independent terms and this can be written:
\bea
\label{fish}
& & \left.
\frac{1}{2} \includegraphics[bburx=500, bbury=430, bbllx=210,
bblly=340, width=4.3cm]{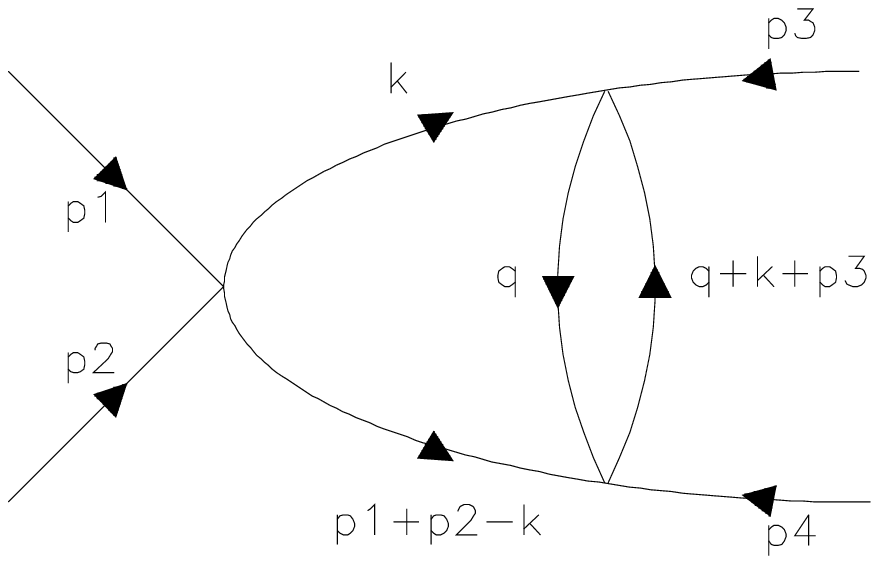} \right|_{\textrm{div NP}} 
~ = \nn \\
& = & \frac{\l^3}{27} \int \frac{\db^4 k ~\db^4 q}{(k^2 + m^2)
((p_1+p_2 -k)^2 + m^2)(q^2 + m^2)((k+q+p_3)^2 +m^2)} ~ \times \nn \\
& \times &\frac{1}{4} \Bigg\{
\cos{\left(\frac{p_1\te p_2 - p_3\te p_4}{2} + k\te (p_1+p_2)\right)}+  
\cos{\left(\frac{p_1\te p_2 + p_3\te p_4}{2} - k\te (p_1+p_2)\right)}+
\nn \\*
& & + \; \cos{\left(\frac{p_1\te p_2 - p_3\te p_4}{2} + k\te
p_1\right)} + \cos{\left(\frac{p_1\te p_2 + p_3\te p_4}{2} - k\te
p_2\right)} + \nn \\*
& & + \;\cos{\left(\frac{p_1\te p_2 - p_3\te p_4}{2} + k\te (p_1+p_3)
\right)}
+\cos{\left(\frac{p_1\te p_2 + p_3\te p_4}{2} + k\te (p_1+p_4)
\right)} \nn \\*
& & + \; 2 \c{p_1}{p_2} \Bigg[\cos{\left(\frac{p_3\te p_4}{2} -k\te p_3
\right)} + \cos{\left(\frac{p_3\te p_4}{2} + k\te p_4\right)}\Bigg]
\Bigg\} \, . \nn \\*
\eea

The factor in front of the fish diagram is $\frac{1}{2}$ because as
explained at the beginning of this section we are considering only
five permutations instead of eleven and we are going to symmetrize the
result with respect to $p_3$ and $p_4$ in the end. So the contribution
of the fish diagram should be written:
\bea
\label{dvnp2}
& & \left. \frac{1}{2} \includegraphics[bburx=500, bbury=430, bbllx=210,
bblly=340, width=3.5cm]{Fig1/fvert2l3.eps} \right|_{\textrm{div NP}} 
~= \nn \\
& = & \frac{\l^3}{27} \int \frac{\db^4 k ~\db^4 q}{(k^2 + m^2)
((p_1+p_2 -k)^2 + m^2)(q^2 + m^2)((k+q+p_3)^2 +m^2)}  \nn \\
& \times &\frac{1}{4} \Bigg\{\frac{1}{2}\bigg[
\cos{\left(\frac{p_1\te p_2 - p_3\te p_4}{2} + k\te (p_1+p_2)\right)}+  
\nn \\*
& & \quad + \;\cos{\left(\frac{p_1\te p_2 + p_3\te p_4}{2} - k\te
(p_1+p_2)\right)}+ p_3~\leftrightarrow~p_4\bigg] + \nn \\*
& & + \; \frac{1}{2} \left[
\cos{\left(\frac{p_1\te p_2 - p_3\te p_4}{2} + k\te p_1\right)} +
\cos{\left(\frac{p_1\te p_2 + p_3\te p_4}{2} - k\te p_2\right)} +
p_3~\leftrightarrow~p_4\right] + \nn \\*
& & + \; \cos{\left(\frac{p_1\te p_2 - p_3\te p_4}{2} + k\te
(p_1+p_3)\right)}+ \cos{\left(\frac{p_1\te p_2 + p_3\te p_4}{2} + k\te
(p_1+p_4)\right)}+ \nn \\*
& & + \; 2 \c{p_1}{p_2} \Bigg[\cos{\left(\frac{p_3\te p_4}{2} -k\te
p_3\right)} + \cos{\left(\frac{p_3\te p_4}{2} + k\te p_4\right)}\Bigg]
\Bigg\} = \nn \\
& = & \frac{\l^3}{27}
\int \frac{\db^4 k ~\db^4 q}{(k^2 + m^2) ((p_1+p_2 -k)^2 + m^2)(q^2 +
m^2) ((k+q+p_3)^2 +m^2)}\times \nn \\
& & \nn \\
& \times & \frac{1}{4}\Bigg\{ 2\c{p_1}{p_2}\c{p_3}{p_4} \cos{k\te (p_1
+ p_2)} 
+\c{p_3}{p_4} \Bigg[\cos{\left(\frac{p_1\te p_2}{2} +k\te p_1\right)}
+\nn \\*
& & + \;  \cos{\left(\frac{p_1\te p_2}{2} - k\te p_2\right)}\Bigg]
+2 \c{p_1}{p_2} \Bigg[\cos{\left(\frac{p_3\te p_4}{2} -k\te p_3\right)}
+ \cos{\left(\frac{p_3\te p_4}{2} + k\te p_4\right)}\Bigg] ~~\nn \\*
& & + \; \cos{\left(\frac{p_1\te p_2 - p_3\te p_4}{2} + k\te
(p_1+p_3)\right)} + \cos{\left(\frac{p_1\te p_2 + p_3\te p_4}{2} +
k\te (p_1+p_4)\right)} \Bigg\} \, ,
\eea
and now we have truly only 5 more permutations.
Up to now we have just discussed the two loop diagrams which have a divergent
nonplanar part. Some more divergences which can be classified as nonplanar
ones come from the counterterm denoted by (F) in (\ref{g42}). These terms can
be written as follows:
\bdm
\left. \frac{1}{2}\includegraphics[bburx=590, bbury=480,
bbllx=250, bblly=370, width=3.3cm]{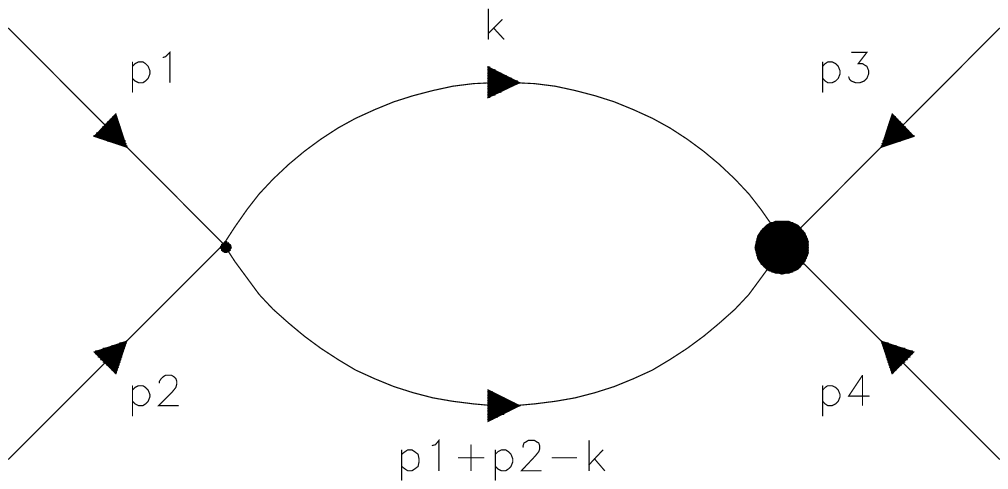}
\right|_{\textrm{nonplanar}} = - \frac{\l^3}{27} \int \frac{\db^4 k
~\db^4 q}{((p_1+p_2-k)^2 + m^2)(k^2 + m^2)(q^2 + m^2)^2}~ \times 
\edm
\bea
\label{dvnp3}
\times~ \Bigg\{\c{p_1}{p_2}\c{p_3}{p_4} \cos{k\te (p_1+p_2)} & + &
\frac{1}{2} \c{p_1}{p_2}\bigg[\cos{\left(\frac{p_3 \te p_4}{2} - 
k\te p_3 \right)} + \nn \\*
+ \cos{\left(\frac{p_3 \te p_4}{2} + k\te p_4 \right)}
\bigg] & + &  
\frac{1}{4}\cos{\left(\frac{p_1\te p_2 + p_3\te p_4}{2} + k\te
(p_1+p_4) \right)} + \nn \\* 
+ \frac{1}{4}\cos{\left(\frac{p_1\te p_2 + p_3\te p_4}{2}
- k\te (p_1+p_3) \right)} & + &
\frac{1}{2} \c{p_3}{p_4}\bigg[\cos{\left(\frac{p_1 \te 
p_2}{2} + k\te p_1 \right)} + \nn \\*
& & \hspace{1.85cm}
+ \cos{\left(\frac{p_1 \te p_2}{2} - k\te p_2 \right)} \bigg]
\Bigg\} \, . \nn \\*
\eea

%

Now we have to combine the results found for the divergent nonplanar pieces in
(\ref{dvnp1}), (\ref{dvnp2}) and (\ref{dvnp3}) and show that at the end of the
day we remain only with a finite piece. The quantity we are dealing with is:
\bea
\label{div2l}
\left[\frac{1}{4}\left(\includegraphics[bburx=500, bbury=430, 
bbllx=270, bblly=340, width=2.7cm]{Fig1/fvert2l2.eps} ~+~
2~~\textrm{perm} \right) \right. & + &\left. 
\frac{1}{2} \left(\includegraphics[bburx=510, bbury=450, bbllx=200,
bblly=340, width=2.7cm]{Fig1/fvert2l3.eps} ~+~
5~~\textrm{perm} \right) \right]_{\textrm{div. NP}}~+ \nn \\*
& + &  \left[\frac{1}{2}\left(\includegraphics[bburx=590, bbury=480,
bbllx=230, bblly=370, width=2.5cm]{Fig1/fvertct2.eps} +~
5~~\textrm{perm} \right)\right]_{\textrm{NP}} \, ,\nn \\
\eea
and in order to show that this is finite we will check separately each
independent cosine 
combination. Let us start with $\c{p_1}{p_2}\c{p_3}{p_4} \cos{k\te(p_1+p_2)}$.
This term can be found in all the three pieces of (\ref{div2l}). The last two
graphs come together with five permutations, but due to the conservation of
momenta i.e. $p_1+p_2~=~-(p_3+p_4)$ only two out of the five permutations are
independent:
\begin{eqnarray}
& &   \frac{1}{2}\Bigg[\c{p_1}{p_2}\c{p_3}{p_4}\cos{k\te(p_1+p_2)} +
  \textrm{5 permutations} \Bigg] = \nn \\
 & = & \c{p_1}{p_2}\c{p_3}{p_4}\cos{k\te(p_1+p_2)} +
  \c{p_1}{p_3}\c{p_2}{p_4}\cos{k\te(p_1+p_3)} \nn \\*
 & & + \c{p_1}{p_4}\c{p_2}{p_3}\cos{k\te(p_1+p_4)} \nn \\
& = & \Bigg[\c{p_1}{p_2}\c{p_3}{p_4}\cos{k\te(p_1+p_2)} +
  \textrm{2 permutations} \Bigg]\, .
\end{eqnarray}
Now it can be seen that this term can be written as:
\bea
& & \frac{\l^3}{27} \Bigg\{ \int \frac{\db^4 k ~\db^4
q}{(k^2 + m^2) ((p_1+p_2 -k)^2 + m^2)(q^2 + m^2)}
\c{p_1}{p_2}\c{p_3}{p_4} \cos{k\te(p_1+p_2)}~ \nn \\*
& & \qquad\times \left[\frac{1}{(q+k+p_3)^2 +m^2} +
\frac{1}{(p_1+p_2-q)^2 + m^2} - \frac{2}{q^2 + m^2}\right] 
+\textrm{2 perm} \Bigg\}\, , \nn 
\eea
and this is obviously finite since the integration over $k$ is regulated by
the cosine factors, while for $q$ the difference of the three propagators
together with the overall $\frac{1}{q^2+ m^2}$ is just enough to guarantee the
convergence of the integral.

The next cosine combination we are going to consider is:
\begin{equation}
  \frac{1}{4}\left[\cos{\left(\frac{p_1\te p_2 + p_3\te
   p_4}{2} + k\te (p_1+p_4)\right)} + \cos{\left(\frac{p_1\te p_2 -
   p_3\te p_4}{2} + k\te (p_1+p_3)\right)} \right]\, .
\end{equation}
This can only be found in the fish diagram and in the counterterm and the
total contribution is:
\begin{eqnarray}
\frac{\l^3}{27} \Bigg\{ \int \frac{\db^4 k ~\db^4 q}{(k^2 + m^2)
((p_1+p_2 -k)^2 + m^2)(q^2 + m^2)} \left[\frac{1}{(q+k+p_3)^2 + m^2} -
\frac{1}{q^2+m^2}\right]  \nn \\*
\times\frac{1}{4}\left[\cos{\left(\frac{p_1\te p_2 + p_3\te
p_4}{2} + k\te (p_1+p_4)\right)} + \cos{\left(\frac{p_1\te p_2 -
p_3\te p_4}{2} + k\te (p_1+p_3)\right)} \right] 
+ \textrm{5 perm} \Bigg\} \nn\\
\end{eqnarray}
The same argument as before can be applied to show that this is finite.

Before going on to the last combination of cosines, let us note some technical
fact regarding the permutations together with which various terms appear.
\bea
& & \Bigg\{ \int \frac{\db^4 k ~\db^4 q}{(k^2 + m^2)
((p_1+p_2 -k)^2 + m^2)(q^2 + m^2)} \cdot \c{p_1}{p_2}
\hspace{3.5cm}\nn \\
& & \qquad \times \left[\cos{\left( \frac{p_3\te p_4}{2}
- k\te p_3 \right)} +  \cos{\left( \frac{p_3\te p_4}{2} + k\te p_4
\right)} \right] + \textrm{5 perm} \Bigg\} \nn \\
& = & \Bigg\{ \int \frac{\db^4 k ~\db^4 q}{(k^2 + m^2)
((p_1+p_2 -k)^2 + m^2)(q^2 + m^2)} \c{p_1}{p_2} \nn \\
& & \times \left[\cos{\left(\frac{p_3\te p_4}{2}
- k\te p_3 \right)} +  \cos{\left( \frac{p_3\te p_4}{2} + k\te p_4
\right)} \right] \nn \\
& + & \int \frac{\db^4 k ~\db^4 q}{(k^2 + m^2)
((p_3+p_4 -k)^2 + m^2)(q^2 + m^2)} \cdot\c{p_3}{p_4}  \nn \\
& & \times\left[\cos{\left(\frac{p_1\te p_2}{2}
- k\te p_1 \right)} +  \cos{\left( \frac{p_1\te p_2}{2} + k\te p_2
\right)} \right] + \textrm{2 perm} \Bigg\}
\eea
Using the conservation of momenta $(p_3+p_4-k)^2$ can be replaced by
$(p_1+p_2+k)^2$. Now by changing $k$ with $-k$ the above result can be
written:
\bea
& &  \Bigg\{ \int \frac{\db^4 k ~\db^4 q}{(k^2 + m^2)
((p_1+p_2 -k)^2 + m^2)(q^2 + m^2)} \nn \\
& & \quad\Bigg[\c{p_3}{p_4}\left( \cos{\left(\frac{p_1\te p_2}{2} + k\te p_1
    \right)}  +\cos{\left(\frac{p_1\te p_2}{2} - k\te p_2 \right)} \right) +
\nn \\ 
& & \qquad+\c{p_1}{p_2}\left(\cos{\left( \frac{p_3\te p_4}{2} -
k\te p_3 \right)} +\cos{\left( \frac{p_3\te p_4}{2} + k\te p_4
\right)}\right) \Bigg] + \textrm{2 perm} \Bigg\} \nn \\
\eea
With these considerations the remaining terms in (\ref{div2l}) can be written:
\bea
& & \frac{\l^3}{27} \Bigg\{\int \frac{\db^4 k ~\db^4 q}{(k^2 + m^2)
((p_1+p_2 -k)^2 + m^2)(q^2 + m^2)}  \nn \\*
& & \qquad \times \frac{1}{4}\Bigg[
\c{p_3}{p_4}\left( \cos{\left(\frac{p_1\te p_2}{2} + k\te p_1 \right)} 
+\cos{\left(\frac{p_1\te p_2}{2} - k\te p_2 \right)} \right) + 
\nn \\*
& & \quad\qquad+\c{p_1}{p_2}\left(\cos{\left( \frac{p_3\te p_4}{2} -
k\te p_3 \right)} +\cos{\left( \frac{p_3\te p_4}{2} + k\te p_4
\right)}\right) \Bigg]  \nn \\*
& & \hspace{3.7cm}\times \left[\frac{1}{(p_1+p_2-q)^2} +
\frac{3}{(q+k+p_3)^2 + m^2} -
\frac{4}{q^2 +m^2}\right] + \textrm{2 perm} \Bigg\}\, . \nn \\*
\eea
As before, we notice that this term is obviously finite due to
the right combination of the propagators in the last square bracket.
The interesting phenomenon which can be observed in these last formulas is
that the terms coming from the nonplanar parts do not really cancel, but they
combine in such a way that the final result remains finite.

Summarizing, the divergences coming from the nonplanar part of diagrams (G)
and (H) in eq. (\ref{g42}) are canceled against the nonplanar part of
the counterterm (F). With this the proof of renormalizability of the
\nc{e} $\Phi^4$ theory up to two loops is complete.




\section{Conclusions and remarks}

In this work we have studied the field theories written on the \nc{e}
Moyal plane (\nc{e} field theories). These field theories are obtained
by replacing the usual product of fields by the star product. First we
discussed some issues of these theories at classical level, then using
the usual methods we quantized the theory. We discussed both canonical
and path integral methods. Because of the star product properties, the
quadratic part of the action is not changed and hence only in the
interaction part one can trace the \nc{ity}. Extending this fact to
the quantum level, we assumed that the Fock space for a commutative
field theory and its \nc{e} version are the same. In the path
integral formulation this means that the measure for the commutative
and \nc{e} theories should be the same, and we support this by
formulating our theory in the momentum space. We should also remind
that in this work we mainly restrict ourselves to the \nc{e} space;
the issue of \nc{e} space-time because of having some problems with unitarity
and causality, seems to be more involved and subtle.

Having developed the necessary ingredients, we worked out the one and
two loops two and four point functions for a \nc{e} $\Phi^4$ theory in
4 dimensions, and presented all the detailed (and maybe tedious)
calculations. 
The important point to be noticed in the \nc{e} cases is that although the
counter-terms arise from the planar parts of the diagrams and hence they have
the 
same divergence structure as the commutative case, they appear to have
different 
numeric factors. Furthermore, these counter-terms are not ${\theta}$ dependent.
The latter means that the $\theta\to 0$ limit is not a smooth limit, and
considering the quantum (loop) corrections we are not recovering the usual commutative
field theory in this limit.  

The other point we should address here is that because of this IR/UV mixing it
is not yet clear that the usual Wilsonian renormalization and
renormalization group arguments 
should work all the same in the \nc{e} case as well.

The result which we want
to emphasize on here is that the noncommutativity parameter, $\theta$, is
not receiving any loop corrections, even at two loops and we expect this
result to be an exact one, 
i.e. $\theta$ is exact, without any quantum corrections at all loops. The
other interesting question 
which we did not address here is the problem of gauge fields and gauge fixing,
and extending the 
present work to gauge theories + fermions, which we hope to come back to in
later works.

\vskip 3cm
\newpage
\begin{center}
\textbf{Acknowledgments}
\end{center}

One of us, A.M. ,  would like to thank all the Professors of HEP Diploma Course
and Ms. Concetta Mosca for her endless help throughout the year. 
We would also thank T. Krajewski for reading the draft and remarks.
This work was partly supported by the EC contract no. ERBFMRX-CT 96-0090.





\appendix
\section{Functional derivatives in star product formalism}
\label{fd}

In this appendix we are going to show what we mean by taking functional
derivatives of terms which contain star products. First we are going to
adopt the usual definition for the functional derivative, i.e.
\bea
S[\Phi + \delta \Phi] - S[\Phi] & \equiv & \int d^4 x ~ \frac{\delta
S[\Phi]}{\delta \Phi(x)} \; \delta \Phi(x) \, .
\eea

Let us apply this definition to the $\Phi^4$ theory:
\bea
S_{int}[\Phi + \delta \Phi] - S_{int}[\Phi] & =  &\frac{\l}{4!}\bigg\{
\int d^4 x ~\Big[ \big((\Phi+\delta \Phi) \star \Phi \star \Phi \star
\Phi\big)(x) + \big( \Phi \star (\Phi+\delta \Phi) \star \Phi \star
\Phi \big)(x)\nn \\  
& &~\: + \big(\Phi \star \Phi \star (\Phi+\delta \Phi) \star \Phi\big)
+ \big(\Phi \star \Phi \star \Phi \star (\Phi+\delta \Phi)\big)(x)
\Big] \nn \\
& &~\: - \int d^4 x ~\big(\Phi \star \Phi \star \Phi \star
\Phi\big)(x) \bigg\} \nn \\ 
& = & \int d^4 x ~\big(\delta \Phi \star \Phi \star \Phi \star \Phi
\big)(x) + \int d^4 x ~\big(\Phi \star \delta \Phi \star \Phi \star
\Phi \big)(x)\nn \\ 
& & ~+ \int d^4 x ~\big(\Phi \star \Phi \star \delta \Phi \star \Phi
\big)(x) + \int d^4 x ~\big(\Phi \star \Phi \star \Phi \star \delta
\Phi\big)(x) \, .\nn \\ 
\eea
Making use of the cyclic property (\ref{pr4}) and of the associativity
of star product (\ref{pr3}) we can write:
\bea
\int d^4x~ \frac{\delta S_{int}[\Phi]}{\delta \Phi(x)} \; \delta
\Phi(x) & = &  \frac{\l}{3!} ~ \int d^4x \;\Big[ \big(\Phi \star \Phi
\star \Phi\big) \star \delta \Phi \Big](x)\nn \\
& = & \frac{\l}{3!} ~ \int d^4x \;  \big(\Phi \star \Phi \star
\Phi\big)(x) \cdot \delta \Phi(x)
\eea
so that we can identify
\bea
\label{d1sint}
\frac{\delta S_{int}[\Phi]}{\delta \Phi(x)} & = & \frac{\l}{3!}~
\big(\Phi \star \Phi \star \Phi\big)(x) \, .
\eea

In the path integral formalism we need to know all possible functional
derivatives of our action with respect to the fields. In the following we are
going to show explicitly the right way to compute these functional derivatives
for the case of the scalar field theory with $\Phi^4$ interaction.
First let us note that due to the conjugation property of the star product
(\ref{pr6}) the r.h.s of (\ref{d1sint}) is real if the field $\Phi$ is real.
However, the next order functional derivatives do not enjoy this property
anymore, and one should make the result to be real explicitly.

\bea
\frac{\delta^2 S_{int}}{\delta \Phi(x_1) \delta \Phi(x_2)} & = &
-\frac{\l}{6} \; Re\Bigg\{ \st{\xi}{\eta} \str{\alpha}{\beta} ~ \times
\nn \\*
& & \times \bigg[ \delta(x_1 +\xi -x_2)\; \Phi(x_1 + \eta +\alpha)
\;\Phi(x_1 + \eta + \beta) + \nn \\* 
& & \quad + \Phi(x_1 + \xi)\; \delta(x_1 +\eta + \alpha -x_2) \;
\Phi(x_1 + \eta + \beta)  + \nn \\*
& & \qquad + \Phi(x_1 + \xi) \; \Phi(x_1 + \eta +\alpha) \;\delta
(x_1 + \eta + \beta - x_2) \bigg] \Bigg\} \, , \\
\label{d3s}
\frac{\delta^3 S_{int}}{\delta \Phi(x_1) \delta \Phi(x_2) \delta
\Phi(x_3)} & = & -\frac{\l}{6}\; Re~ \Bigg\{\st{\xi}{\eta}
\str{\alpha}{\beta} ~\times \nn \\*
& & \times \bigg[ \delta(x_1 +\xi -x_2) \delta(x_1 + \eta +\alpha
-x_3)\;\Phi(x_1 + \eta + \beta) + \nn \\*
& & \phantom{+} + \delta(x_1 +\xi -x_2)\; \Phi(x_1 + \eta +\alpha) \;
\delta (x_1 + \eta + \beta -x_3 ) \nn \\*
& & \quad + \delta(x_1 + \xi - x_3) \; \delta(x_1 +\eta + \alpha -x_2)
\; \Phi(x_1 +\eta + \beta) \nn \\*
& & \quad + \Phi(x_1 + \xi)\; \delta(x_1 +\eta + \alpha -x_2) \;
\delta(x_1 +\eta + \beta - x_3) \nn \\*
& & \quad + \delta(x_1 + \xi - x_3) \; \Phi(x_1 + \eta +\alpha)
\;\delta  (x_1 + \eta + \beta - x_2) \nn \\*
& & \qquad + \Phi(x_1 + \xi) \; \delta(x_1 + \eta +\alpha -x_3) 
\delta(x_1 + \eta + \beta - x_2) \; \bigg] \Bigg\}  \, , \nn \\
\eea 
\bea
\label{d4s}
\frac{\delta^4 S_{int}}{\delta \Phi(x_1) \delta \Phi(x_2) \delta \Phi(x_3)
\Phi(x_4)} & = & -\frac{\l}{6}\;Re~ \Bigg\{ \st{\xi}{\eta}
\str{\alpha}{\beta} ~\times \nn \\*
& & \times \bigg[ \delta(x_1 +\xi -x_2) \delta(x_1 + \eta +\alpha
-x_3)\; \delta(x_1 + \eta + \beta -x_4) + \nn \\*
& & \quad + \delta(x_1 +\xi -x_2)\; \delta(x_1 + \eta +\alpha
-x_4) \; \delta (x_1 + \eta + \beta -x_3 ) \nn \\*
& & \quad + \delta(x_1 + \xi - x_3) \; \delta(x_1 +\eta + \alpha -x_2) \;
\delta(x_1 +\eta + \beta - x_4) \nn \\*
& & \quad + \delta(x_1 + \xi -x_4)\; \delta(x_1 +\eta + \alpha
-x_2) \; \delta(x_1 +\eta + \beta - x_3) \nn \\*
& & \quad+ \delta(x_1 + \xi - x_3) \; \delta(x_1 + \eta +\alpha -x_4)
\;\delta (x_1 + \eta + \beta - x_2) \nn \\*
& & \qquad + \delta(x_1 + \xi -x_4) \; \delta(x_1 + \eta +\alpha
-x_3)  \delta(x_1 + \eta + \beta - x_2) \bigg] \Bigg\} \, . \nn \\
\eea 

The above expression looks simpler in momentum space:
\bea
\label{d4sms}
\frac{\delta^4 S_{int}}{\delta \Phi^4} & = & - \frac{\l}{6} Re~
\Bigg\{ \int \db^4 k_2 \db^4 k_3 \db^4 k_4 e^{i k_2(x_1-x_2)} e^{i
k_3(x_1-x_3)} e^{i k_4(x_1-x_4)} 
\st{\xi}{\eta} \str{\alpha}{\beta} \nn \\*
& & \Big[e^{i k_2 \xi}\; e^{i (k_3+k_4) \eta} \; e^{i k_3 \alpha} e^{i
k_4 \beta} + e^{i k_2 \xi}\; e^{i (k_3+k_4) \eta} \; e^{i k_4 \alpha}
e^{i k_3 \beta} + 
e^{i k_3 \xi}\; e^{i (k_2 + k_4) \eta} \; e^{i k_2 \alpha} e^{i k_4
\beta} \nn \\*
& & + e^{i k_4 \xi}\; e^{i (k_2+k_3) \eta} \; e^{i k_2 \alpha} e^{i
k_3 \beta} + 
e^{i k_3 \xi}\; e^{i (k_2+k_4) \eta} \; e^{i k_4 \alpha} e^{i k_2
\beta} + e^{i k_4 \xi}\; e^{i (k_2+k_3) \eta} \; e^{i k_3 \alpha} e^{i
k_2 \beta}  \Big]  \Bigg\} \nn \\
\nn \\
& = & - \frac{\l}{6} Re~ \Bigg\{\int \db^4 k_2 \db^4 k_3 \db^4 k_4
e^{i k_2(x_1-x_2)} e^{i k_3(x_1-x_3)} e^{i k_4(x_1-x_4)} \Big[
\kstm{k_2}{(k_3+k_4)} \; \kstm{k_3}{k_4}  \nn \\*
& & ~ + \kstm{k_2}{(k_3+k_4)} \; \kstm{k_4}{k_3} +
\kstm{k_3}{(k_2+k_4)} \; \kstm{k_2}{k_4} + \kstm{k_3}{(k_2+k_4)} \;
\kstm{k_4}{k_2} \nn \\*
& & \quad + \kstm{k_4}{(k_2+k_3)} \; \kstm{k_2}{k_3} +
\kstm{k_4}{(k_2+k_3)} \; \kstm{k_3}{k_2} \Big] \Bigg\} \nn \\
\nn \\
& = & -\frac{\l}{3} \int \db^4k_1 \; \db^4 k_2 \; \db^4 k_3 \; \db^4
k_4  e^{-i k_1 x_1 - i k_2 x_2 - i k_3 x_3 - i k_4 x_4} ~ (2 \pi)^4 \;
\delta (k_1 +k_2+k_3 +k_4) \nn \\*
& & \times \Big[\c{k_1}{k_2}\; \c{k_3}{k_4} + \c{k_1}{k_3}\;
\c{k_2}{k_4} + \c{k_1}{k_4}\; \c{k_2}{k_3} \Big] \, .
\eea

\section{Phase factors associated with various diagrams}
\label{pf}

In this section we are going to compute explicitly the phase factors
associated with the diagrams we encounter in the two loop expansion of the
four point function (\ref{g42})

Let us first introduce a notation for the factors associated with the one
loop diagram:
\bea
\label{v1l}
\frac{1}{2}\includegraphics[bburx=590, bbury=480, bbllx=230,
bblly=370, width=4cm]{Fig1/vert1l.eps} & \propto & \mathcal{P}(p_1
\ldots p_4,k,\te)\, ,~~\textrm{where}
\eea
\bea
\label{cf1l}
\mathcal{P}(p_1 \ldots p_4,k,\te) & = & 
\frac{\l^2}{9} \c{p_1}{p_2} \c{p_3}{p_4}\Bigg[1+\cos{k\te
(p_1+p_2)}\Bigg] ~+ \nn \\
& + & \frac{\l^2}{18} \c{p_3}{p_4}\Bigg[\cos{\left(\frac{p_1\te
p_2}{2}+ k \te p_1 \right)} + \cos{\left(\frac{p_1\te p_2}{2} - k \te
p_2 \right)} \Bigg] ~+ \nonumber \\
& + & \frac{\l^2}{18} \c{p_1}{p_2}\Bigg[\cos{\left(\frac{p_3\te
p_4}{2}- k \te p_3 \right)} + \cos{\left(\frac{p_3\te p_4}{2} + k \te
p_4 \right)} \Bigg] ~+ \nonumber \\
& & \hspace{-2cm} + \frac{\l^2}{36}\Bigg[\cos\left(\frac{p_1\te p_2
-p_3\te p_4}{2} + 
k\te(p_1+p_3)\right) + \cos\left(\frac{p_1\te p_2 +p_3\te p_4}{2}
+ k\te(p_1+p_4)\right)\Bigg] \, . \nonumber \\*
\eea 

If we want to find the planar part of this graph we should just drop the terms
which involve internal momenta and taking into account the two permutations
together with which this diagram appears in the loop expansion we conclude:
\begin{eqnarray}
Planar[\mathcal{P}(p_1 \ldots p_4,k,\te)] = \frac{\l^2}{9} \left(
\c{p_1}{p_2} \c{p_3}{p_4} + \textrm{2 perm}\right)\, ,
\end{eqnarray}
so that the momentum dependent factor is just the same as the usual factor we
added for the \nc{e} vertex (\ref{factv}). 
We shall now compute the phase factors associated to each diagram in
the expansion of $\G^{(4)}$ in (\ref{g42}) leaving apart for the
moment the overall factor $\frac{1}{27}$. 

\bea
\label{cf2l}
(G) & \propto & \Bigg[ \c{(k-q)}{(p_1+p_2)} +
\cos{\left(\frac{(k+q)\te (p_1+p_2)}{2} -k\te q\right)} +
\c{(k+q)}{(p_1+p_2)}\Bigg]\times  \nn  \\
& \times & \Bigg[ \cos{\frac{p_1\te p_2 +k\te (p_1+p_2)}{2}} +~ 
\cos{\frac{p_1\te p_2 -k\te (p_1+p_2)}{2}} +~
\cos{\frac{p_1\te p_2 +k\te (p_1-p_2)}{2}} \Bigg]\times  \nn\\
& \times & \Bigg[\cos{\frac{p_3\te p_4 +q\te (p_3+p_4)}{2}} +~ 
\cos{\frac{p_3\te p_4 - q\te (p_3+p_4)}{2}} +~ \cos{\frac{p_3\te p_4
-q\te (p_3 - p_4)}{2}} \Bigg] \, .\nn \\
\eea

At two-loops we have two independent
internal momenta to integrate over so, as explained in section \ref{sg22} all
the terms which contain cosine factors and involve both of these momenta
will remain finite after integrations.
The second term in the first factor from eq. (\ref{cf2l}) contains
a $k\te q$ in the argument of the cosine which cannot be found
anywhere else. So after expanding and transforming the cosine
products into sums of cosines this term cannot disappear. In what
follows we consider only terms from which either $k$ or
$q$ (or both) disappear. These terms come from:

\bea
\frac{1}{2}\Bigg[ \cos{\frac{p_3\te p_4 +q\te (p_3+p_4)}{2}} + 
\cos{\frac{p_3\te p_4 - q\te (p_3+p_4)}{2}}+\cos{\frac{p_3\te
p_4 +q\te (p_3 - p_4)}{2}} \Bigg] \times \nn \\
\times \Bigg[\cos{\left(\frac{p_1\te p_2 -q\te
(p_1+p_2)}{2}+k\te(p_1+p_2) \right)}+\cos{\left(\frac{p_1\te 
p_2 + q\te (p_1+p_2)}{2}\right)} + \nn \\
+ \cos{\left(\frac{p_1\te p_2 + q\te (p_1+p_2)}{2} - k\te(p_1+p_2)
\right)}+ 
\cos{\left(\frac{p_1\te p_2 - q\te (p_1+p_2)}{2}\right)} + \nn \\
+ \cos{\left(\frac{p_1\te p_2 + q\te (p_1+p_2)}{2} + k\te (p_1+p_2) 
\right)}+\cos{\left(\frac{p_1\te p_2 - q\te
(p_1+p_2)}{2}\right)} +  \nn \\
+\cos{\left(\frac{p_1\te p_2 - q\te (p_1+p_2)}{2} - k\te(p_1+p_2)
\right)}+
\cos{\left(\frac{p_1\te p_2 + q\te (p_1+p_2)}{2}\right)}  +
\nn \\
+\cos{\left(\frac{p_1\te p_2 - q\te (p_1+p_2)}{2} + k\te p_1
\right)}+\cos{\left(\frac{p_1\te p_2 + q\te (p_1+p_2)}{2} - k\te
p_2 \right)} + \nn \\
+ \cos{\left(\frac{p_1\te p_2 + q \te (p_1+p_2)}{2} + k \te p_1 
\right)}+\cos{\left(\frac{p_1\te p_2 -q \te (p_1+p_2)}{2} - k\te
p_2\right)} \Bigg]  \nn \, . \, \\
\eea

\paragraph{The planar part} Using overall momentum conservation we can
extract the planar part:
\bea
\textrm{Planar}[(G)] & = &
2\cdot\frac{1}{2}\left[\cos{\frac{p_3\te p_4 +q\te (p_3+p_4)}{2}} + 
\cos{\frac{p_3\te p_4  - q\te (p_3+p_4)}{2}}\right] \times \nn \\*
& & ~~~ \left. \times \left[\cos{\frac{p_1\te p_2 +q\te(p_1+p_2)}{2}}
+ \cos{\frac{p_1\te p_2  - q\te (p_1+p_2)}{2}}
\right]\right|_{\textrm{Planar}}~ = \nn \\*
& = & \cos{\frac{p_1\te p_2 + p_3\te p_4}{2}} +
\cos{\frac{p_1\te p_2 - p_3\te p_4}{2}} \nn \\*
& = & 2\cdot \c{p_1}{p_2}\cdot\c{p_3}{p_4} \, .
\eea
The ``2 permutations'' take care of the other combinations of external
momenta, and so
\bea
\frac{1}{4}\left[\includegraphics[bburx=500, bbury=400, bbllx=270, 
bblly=340, width=2.5cm]{Fig1/fvert2l2.eps}
+ 2~\textrm{permutations} \right]_{\textrm{planar}} & = & \frac{1}{2
\cdot 27}\Bigg[
\c{p_1}{p_2}\cdot\c{p_3}{p_4} + \nn \\*
& + & \c{p_1}{p_3}\cdot\c{p_2}{p_4} + \c{p_1}{p_4}\cdot\c{p_2}{p_3} 
\Bigg] \nn \\*
\eea

\paragraph{Nonplanar q-independent terms}
\bea
\frac{1}{4}\left. \includegraphics[bburx=500, bbury=400, bbllx=270,
bblly=340, width=2.5cm]{Fig1/fvert2l2.eps} \right|_{\textrm{NP I}}
\hspace{-.5cm} & = &
\frac{1}{16} \Bigg[ 2\cos{\left(\frac{p_1\te p_2 - p_3\te
p_4}{2} + k\te (p_1+p_2)\right)} \nn \\ 
& & + \; 2\cos{\left(\frac{p_1\te p_2 + p_3\te p_4}{2} + k\te (p_1+p_2)
\right)} \nn \\
& & + \; 2\cos{\left(\frac{p_1\te p_2 + p_3\te p_4}{2} 
- k\te (p_1+p_2) \right)} \nn \\
& & + \; 2\cos{\left(\frac{p_1\te p_2 - p_3\te p_4}{2} - k\te (p_1+p_2)
\right)} \nn \\ 
& & + \;  2\cos{\left(\frac{p_1\te p_2 - p_3\te p_4}{2} + k\te p_1
\right)} +
2\cos{\left(\frac{p_1\te p_2 + p_3\te p_4}{2} + k\te p_1 
\right)} \nn \\
& & + \;  2\cos{\left(\frac{p_1\te p_2 + p_3\te p_4}{2} - k\te
p_2\right)} + 2\cos{\left(\frac{p_1\te p_2 - p_3\te p_4}{2} - k\te
p_2\right)} \Bigg]\nn \\
& = & \frac{1}{2} \c{p_1}{p_2}\c{p_3}{p_4} \cos{k\te (p_1+p_2)} \nn \\
& & + \; \frac{1}{4}\c{p_3}{p_4} \left[\cos{\left(\frac{p_1\te p_2}{2} +
k\te p_1 \right)} + \cos{\left(\frac{p_1\te p_2}{2} - k\te
p_2\right)}\right] \nn \, .\\
\eea

\paragraph{Nonplanar k-independent terms}

The diagram is perfectly symmetric in $k$ and $-q$ so we can just
replace $k$ by $-q$ to get:
\bea
\frac{1}{4}\left. \includegraphics[bburx=500, bbury=400, bbllx=270,
bblly=330, width=2.5cm]{Fig1/fvert2l2.eps} \right|_{\textrm{NP II}}
& = & 
\frac{1}{2} \c{p_1}{p_2}\c{p_3}{p_4} \cos{q\te (p_1+p_2)} \nn \\
& + & \frac{1}{4}\c{p_1}{p_2} \left[\cos{\left(\frac{p_3\te p_4}{2} - q\te p_3
\right)} + \cos{\left(\frac{p_3\te p_4}{2} + q\te p_4\right)}
\right]\, . \nn \\
\eea

Let us now consider the next term in eq. (\ref{g42}). The phase
factors associated with the vertices are:
\bea
(H) & \propto & \Bigg[ \cos{\frac{p_1\te p_2 +k\te (p_1+p_2)}{2}} + 
\cos{\frac{p_1\te p_2 -k\te (p_1+p_2)}{2}} + \cos{\frac{p_1\te p_2 +
k\te (p_1-p_2)}{2}} \Bigg]\times  \nn\\ 
& \times & \Bigg[\cos{\frac{p_3\te k - q\te p_3 -q\te k}{2}} +
\cos{\frac{p_3 \te k + q\te p_3 + q\te k}{2}} + \cos{\frac{p_3\te k -
q\te p_3 + q\te k}{2}} \Bigg]\times  \nn\\ 
& \times & \Bigg[\cos{\frac{p_3 \te p_4 + k\te p_4 + q\te p_3 + q \te 
k}{2}} + \cos{\frac{p_3 \te p_4 + k\te p_4 - q\te p_3 - q \te k}{2}} +
\nn \\
& & \hspace{5.5cm} 
+\cos{\left(\frac{p_3 \te p_4 + k\te p_4 + q\te p_3 + q \te k}{2} +
q\te p_4  \right)}\Bigg]\, .
\eea

As before the terms containing simultaneously $k$ and $q$ in the
argument of cosine factor give no contribution to the divergent part.
This means that in the product of the last two terms we only have to
consider those
combinations of cosines which do not lead to factors of $k\te q$ in
the argument. It is easy to see that these terms come from:
\bea
& & \hspace{-1cm}\frac{1}{2} 
\Bigg[ 2\cos{\frac{p_3\te p_4 - k\te (p_3 - p_4)}{2}} 
+2 \cos{\frac{p_3\te p_4 + k\te (p_3 + p_4)}{2}} + \cos{\left( \frac{p_3
\te p_4 - k\te (p_3-p_4)}{2}+q\te p_4\right)} \nn \\
& + & \cos{\left(\frac{p_3\te p_4 + k\te (p_3 + p_4)}{2}+q\te p_4 
\right)} + \cos{\left( \frac{p_3\te p_4 + k\te (p_3 + p_4)}{2} - q\te
p_3 \right)} \nn \\
& + & \cos{\left(\frac{p_3\te p_4 + k\te (p_4 - p_3)}{2} + q \te p_3
\right)} + \cos{\left(\frac{p_3\te p_4 + k\te (p_3 + p_4)}{2}+q\te 
(p_3 +p_4)\right)} \Bigg] \times \nn \\
& \times & \Bigg[ \cos{\frac{p_1\te p_2 +k\te (p_1+p_2)}{2}} + 
\cos{\frac{p_1\te p_2 -k\te (p_1+p_2)}{2}} + \cos{\frac{p_1\te p_2 +
k\te (p_1-p_2)}{2}} \Bigg] \, .
\eea

\paragraph{The planar part} 
Proceeding in the same way as before we can write the planar part:
\bea
\cos{\frac{p_3\te p_4 + k\te (p_3+p_4)}{2}} \cdot \Bigg[\cos{\frac{p_1
\te p_2 + k\te (p_1+p_2)}{2}} & + &\cos{\frac{p_1\te p_2 - k\te
(p_1+p_2)}{2}}\Bigg]\Bigg|_{\textrm{planar}} =  \nn \\
=~\frac{1}{2}\Bigg[ \cos{\frac{p_1\te p_2 + p_3\te p_4}{2}}
& + & \cos{\frac{p_1\te p_2 - p_3\te p_4}{2}} \Bigg] \, , \nn \\
\eea
Taking into account all the coefficients (also the $\frac{1}{27}$) we obtain:
\bea
\frac{1}{2} \left. \includegraphics[bburx=510, bbury=430, bbllx=200,
bblly=340, width=3.3cm]{Fig1/fvert2l3.eps} \right|_{\textrm{Planar}} &
\propto & \frac{1}{2\cdot 27}\c{p_1}{p_2}\c{p_3}{p_4} \, .
\eea

\paragraph{Nonplanar q-independent terms:}

\bea
& & \left. \includegraphics[bburx=510, bbury=430, bbllx=200,
bblly=340, width=3.3cm]{Fig1/fvert2l3.eps} \right|_{\textrm{NP I}} 
\propto \nn \\* 
& \propto & \Bigg[\cos{\frac{p_1\te p_2 + k\te (p_1+p_2)}{2}} +
\cos{\frac{p_1\te p_2 - k\te (p_1+p_2)}{2}}
+ \cos{\frac{p_1\te p_2 + k\te (p_1-p_2)}{2}} \Bigg]  \nn \\
& & \times \Bigg[\cos{\frac{p_3\te p_4 + k\te (p_4-p_3)}{2}} 
 + \cos{\frac{p_3\te p_4 + k\te (p_3+p_4)}{2}}\Bigg] 
\Bigg|_{\textrm{nonplanar}}\!\!\! = \nn \\
& = & \frac{1}{2}\Bigg[\cos{\left(\frac{p_1\te p_2 +p_3\te p_4}{2} -
k\te p_3 \right)} +
\cos{\left(\frac{p_1\te p_2 - p_3\te p_4}{2} - k\te p_4\right)}
\nn \\
& & \quad +\cos{\left(\frac{p_1\te p_2 + p_3\te p_4}{2} + k\te p_4
\right)} + \cos{\left(\frac{p_1\te p_2 - p_3\te p_4}{2} + k\te p_3
\right)}
\nn \\
& & \quad +\cos{\left(\frac{p_1\te p_2 + p_3\te p_4}{2} + k\te (p_1 +
p_4) \right)} +
\cos{\left(\frac{p_1\te p_2 - p_3\te p_4}{2} + k\te (p_1 + p_3)
\right)} \nn \\
& & \quad +\cos{\left(\frac{p_1\te p_2 - p_3\te p_4}{2} + k\te (p_1 +
p_2) \right)} +
\cos{\left(\frac{p_1\te p_2 + p_3\te p_4}{2} - k\te (p_1 + p_2)
\right)} \nn \\
& & \quad +\cos{\left(\frac{p_1\te p_2 + p_3\te p_4}{2} - k\te p_2
\right)} + \cos{\left(\frac{p_1\te p_2 - p_3\te p_4}{2} + k\te p_1
\right)} \Bigg] \, .\nn \\
\eea

\paragraph{Nonplanar k-independent terms:}
\bea
& & \left. \includegraphics[bburx=510, bbury=430, bbllx=200,
bblly=340, width=3.3cm]{Fig1/fvert2l3.eps} \right|_{\textrm{NP II}}
\propto  \nn \\
& \propto & \frac{1}{4}\Bigg[\cos{\left(\frac{p_1\te p_2 +p_3\te
p_4}{2}+q\te p_4 \right)} + \cos{\left(\frac{p_1\te p_2 - p_3\te
p_4}{2} - q\te p_4 \right)} \nn \\
& & \quad + \cos{\left(\frac{p_1\te p_2 + p_3\te p_4}{2} + q\te p_3
\right)} + \cos{\left(\frac{p_1\te p_2 - p_3\te p_4}{2} - q\te p_3
\right)} \nn \\
& & \quad +\cos{\left(\frac{p_1\te p_2 + p_3\te p_4}{2} + q\te (p_3 +
p_4) \right)} + \cos{\left(\frac{p_1\te p_2 - p_3\te p_4}{2} - q\te
(p_3 + p_4)\right)} \Bigg] \, .\nn \\
\eea

Summarizing, the main result we found in this appendix is that the planar
parts of the loop diagrams in (\ref{g42}) come in with the same momentum
dependent phase factor, which is exactly (\ref{factv}). In this way we
motivated the claim that we can renormalize the planar parts as in the
commutative case.


\end{document}